\documentclass[final,5p,times,twocolumn]{elsarticle}
\usepackage{booktabs} % For formal tables
\usepackage{subfig}
\usepackage{amssymb}
\usepackage{lipsum}
\usepackage{natbib}
\newcommand{\etal}{\textit{et al}. }
\usepackage{graphicx}
\usepackage{algorithm}
\usepackage{algpseudocode}%
\usepackage{xcolor}
\usepackage{amsmath}
\usepackage{mathtools}
\usepackage{multirow}
\usepackage{graphics}
\usepackage{enumitem}
\usepackage{CJKutf8}
\usepackage{array}
\usepackage{diagbox}
\usepackage{svg} 
\usepackage{amssymb}
\usepackage{comment}
\usepackage{amsmath}
\usepackage{hyperref}
\usepackage[utf8]{inputenc}
\usepackage{ragged2e}
\usepackage{lettrine}

\makeatother
\biboptions{sort&compress}
\journal{arXiv}

\begin{document}

\begin{frontmatter}



\title{Unsupervised Pairwise Learning Optimization Framework for Cross-Corpus EEG-Based Emotion Recognition Based on Prototype Representation}


\author[1]{Guangli Li}
\author[1]{Canbiao Wu}
\author[4,5,*]{Zhen Liang}

\affiliation[1]{organization={School of Biological Science and Medical Engineering, Hunan University of Technology}, city={Zhuzhou}, country={China}}
\affiliation[4]{organization={School of Biomedical Engineering, Health Science Center, Shenzhen University}, city={Shenzhen}, country={China}}
\affiliation[5]{organization={Guangdong Provincial Key Laboratory of Biomedical Measurements and Ultrasound Imaging}, city={Shenzhen}, country={China}}
\affiliation[*]{Address correspondence to: janezliang@szu.edu.cn}

\begin{abstract}
\indent Affective computing is a rapidly developing interdisciplinary research direction in the field of brain-computer interface, the core goal of which is to achieve accurate recognition of emotional states through physiological signals. In recent years, the introduction of deep learning technology has greatly promoted the development of the field of emotion recognition. However, due to physiological differences between subjects and changes in the experimental environment, cross-corpus emotion recognition faces serious challenges, especially for samples near the decision boundary. To solve the above problems, we propose an optimization method based on domain adversarial transfer learning to fine-grained alignment of affective features. In domain adaptive optimization, we propose Local maximum mean discrepancy (Lmmd) and Contrastive domain discrepancy (Cdd) strategies based on pairwise learning, introduce the theory of Reproducing Kernel Hilbert Space (RKHS), and use feature kernel function $\kappa$ to realize the alignment of sample features. In the rule domain adaptive optimization, we further propose a Maximum classifier discrepancy with Pairwise Learning (McdPL) model to maximize classification discrepancy and minimize feature distribution by designing dual adversarial classifiers (Ada and RMS classifiers) to process samples around decision boundaries, and through three-stage adversarial training. During domain adversarial training, the two classifiers also maintain an adversarial relationship, ultimately enabling precise cross-corpus feature alignment. We conducted systematic experimental evaluation of the model using publicly available SEED, SEED-IV and SEED-V databases. The results show that the McdPL model is superior to other baseline models in the cross-corpus emotion recognition task, achieving SOTA performance with average accuracy improvements of 4.76\% and 3.97\%, respectively. Our work provides a promising solution for emotion recognition across databases. The source code is available at \url{https://github.com/WuCB-BCI/Mcd_PL}.
\end{abstract}



\begin{keyword}
%% keywords here, in the form: keyword \sep keyword, up to a maximum of 6 keywords
EEG \sep Emotion Recognition \sep EEG Processing \sep Optimization Strategy \sep Domain Adversarial Learning.



\end{keyword}

\end{frontmatter}

%\tableofcontents


\begin{figure*}
\centering
\subfloat{\includegraphics[width=1\textwidth]{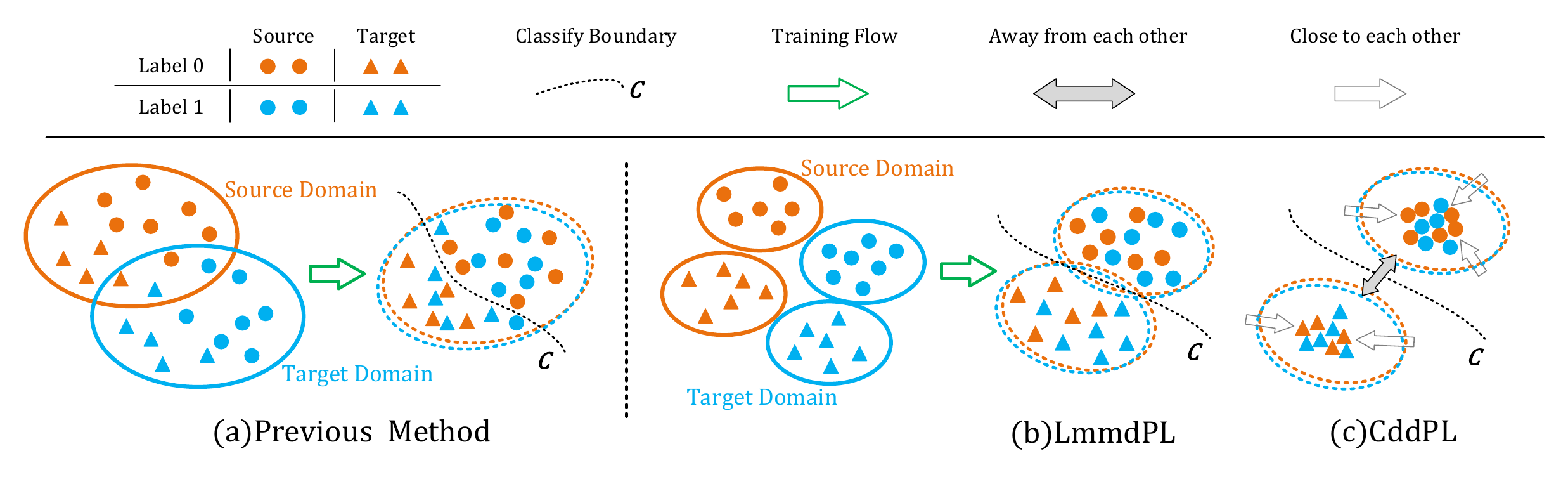}}
\caption{Sample feature alignment schematic diagram of source domain and target domain: (a) sample feature alignment using traditional methods; (b) fine-grained feature alignment in subdomains of the same category in different domains; (c) Within and between classes optimization in opposite directions, similar sample features are close to each other and dissimilar sample features are far away from each other.}
\label{fig:pr_LMMD_cdd_methods}
\end{figure*}
%--------------------------------------------------------------------------

\section{Introduction}
\label{introduction}
\begin{CJK*}{UTF8}{gbsn}{
    \lettrine{A}{ffective} is a psychological state, usually caused by neurophysiological changes, and is one of the basic psychological experiences of human beings \cite{Schacter_Gilbert_Wegner_2011}. It not only affects people's feelings, thinking, and behavior, but also affects people's physical and mental health \cite{2002Affective}. Therefore, how to accurately assess people's emotional state and provide personalized adjustment programs is a difficult problem to be solved. Emotional computing is a rapidly developing interdisciplinary research field, and emotional state recognition is a key issue in the field of Affective computing \cite{Rayatdoost_Soleymani2018}. Traditional emotion recognition mainly relies on non-physiological signals such as facial expressions, voice signals, and body gestures \cite{Kessous_Castellano_Caridakis_2010}, which are highly subjective and the performance is limited. As a kind of physiological signal, electroencephalography (EEG) has the advantages of not being easily camouflaged, excellent real-time performance, objectivity and \textit{etc} \cite{Ye_Zhang_Teng_Zhang_Ni_Liang_2023}\cite{Ye2023AdaptiveSA}. It can provide a more direct and objective clue for understanding and evaluating emotional states, so it has attracted more and more attention from researchers in different fields such as computer science, neuroscience, and signal processing \cite{jung2019utilizing}.
    \\ \indent In recent years, more and more researchers have focused on applying deep learning methods to mitigate individual differences in EEG signals \cite{jayaram2016transfer,Li_Qiu_Shen_Liu_He_2019,Cui_Liu_Zhang_Chen_Wang_Chen_2020,Zhong_Wang_Miao_2022,Gu_Cao_Jolfaei_Xu_Wu_Jung_Lin_2020} and improve feature invariant representation \cite{Ozdenizci_Wang_Koike-Akino_Erdogmus_2019,ozdenizci2020learning,bethge_2022_domain}. Emotion recognition models based on EEG have been widely used and have shown very good performance in emotion recognition tasks in databases.Such as,
    Zhang \etal\cite{Zhang_Yao_2022} introduced both cascade and parallel convolutional recurrent neural network models for precisely identifying human intended movements and instructions, effectively learning the compositional spatio-temporal representations of raw EEG streams.
    Li \etal\cite{Li_Zheng_Wang_Zong_Cui_2022} proposed the R2G-STNN model, which consists of spatial and temporal neural network models with a regional to global hierarchical feature learning process to learn discriminative spatial-temporal EEG features.
    Feng \etal\cite{Feng_Cheng_Zhao_Deng_Zhang_2022} designed a hybrid model called ST-GCLSTM, which comprises a spatial-graph convolutional network (SGCN) module and an attention-enhanced bi-directional Long Short-Term Memory (LSTM) module, which can be used to extract representative spatial-temporal features from multiple EEG channels.
    Yang \etal\cite{10509712_2024} proposed spectral-spatial attention alignment multi-source domain adaptation ($\mathrm{S^2A^2}$-MSD), which constructs domain attention to represent affective cognition attributes in spatial and spectral domains and utilizes domain consistent loss to align them between domains.
    Yan \etal\cite{10510577_2024} proposed the bridge graph attention-based graph convolution network (BGAGCN). It bridges previous graph convolution layers to attention coefficients of the final layer by adaptively combining each graph convolution output based on the graph attention network, thereby enhancing feature distinctiveness.    
    \\ \indent However, 
    \textbf{(1)} Due to the huge discrepancy between different databases \cite{zhou_2023_eeg}, traditional EEG emotion recognition mainly focuses on intra-individual or cross-task. When dealing with EEG emotion recognition tasks in different databases, in addition to dealing with individual discrepancy, it is also necessary to face ‌numerous factors such as the discrepancy in the environment and equipment of EEG collection, resulting in a significant decline in the performance of these emotion recognition methods \cite{Rayatdoost_Soleymani2018}. 
    \textbf{(2)} At present, EEG-emotion experiments are basically induced by video. Subjects may not always respond correctly to emotions due to individual physiological factors, and their emotional changes maybe not be accurately described. This brings unavoidable label noise to the emotional labeling of EEG samples \cite{Jia_Salzmann_Darrell_2010}. The traditional EEG based emotion recognition model is mainly based on pointwise learning and has been successfully applied, but it is highly dependent on accurately labeled EEG data. In contrast, pairwise learning can model the relative association between samples and evaluate the similarity between samples, with less dependence on labels and better robustness and generalization performance \cite{Bao_Niu_Sugiyama_2018,Bao_Shimada_Xu_Sato_Sugiyama_2020,hsu2020deep,Zhao_Zhang_Wu_Moura_Costeira_Gordon_2018}.
    \textbf{(3)} Domain adversarial learning has become a critical important strategy in the field of emotion recognition, significantly reducing the impact of individual discrepancy on recognition performance, and effectively improving the feature representation and generalization ability of the model \cite{ganin2016domain}\cite{Lian_Tao_Liu_Huang_2019}. However, as shown in Fig.\ref{fig:pr_LMMD_cdd_methods}.(a), many domain adversarial learning methods attempt to fully match feature distribution between different domains, but they ignore the decision boundaries of specific tasks between categories, which makes it difficult to accurately identify samples near the boundaries.
    \\ \indent Therefore, in order to solve the ‌aforementioned triple challenges of cross-corpus feature distribution discrepancy, label noise interference and domain confrontation decision boundary ambiguity‌, we innovatively propose a Cross-corpus EEG emotion recognition Transfer model based on \textbf{M}aximum \textbf{c}lassifier \textbf{d}iscrepancy \textbf{P}airwise \textbf{L}earning (McdPL). In addition, we also propose two cross-corpus transfer learning optimization methods (LmmdPL and CddPL), which we will introduce in Sec.\ref{sec:lmmd} and Sec.\ref{sec:cdd}.
    \\ \indent In the McdPL model, we use domain discriminators and feature generators to mitigate the discrepancy in feature distribution between the source domain and the target domain, making it difficult for the discriminator to distinguish whether these features come from the source domain or the target domain. Furthermore, assuming that each emotion has fundamental characteristic attributes‌--named prototype  representation, we extract the prototype representation to explore the potential variables of ‌emotion-category‌ EEG signals, and learn the generalized prototype representation of emotion representation between individuals to enhance the generalization ability of the model. We reformulate emotion recognition as a pairwise learning problem to replace traditional classifiers, thereby ‌significantly‌ reducing the model's dependence on high-precision emotion labels. 
    Therefore, based on the idea of pairwise learning, we propose two classifiers with the same structure but different optimization strategies, and divide the training process into three stages, which are Basic Training, Maximize Classifiers Discrepancy and Minimize Features Distribution. Ultimately, the ‌McdPL model ‌ achieves fine-grained alignment of features in cross-corpus ‌ samples, enhancing the model's ability ‌ to handle samples near the decision boundary.
    Overall, the main contributions of this paper are summarized as follows :
    \begin{itemize}
        \item Based on adaptive optimization of domain adversarial learning, we propose the LmmdPL and CddPL models that focus on the relationship between related subdomains, and the McdPL models that focus on decision boundary Controversy samples. These models effectively ‌ align in the feature space, which enhances model performance.
        \item In the McdPL model, we adopt the unsupervised learning strategy and propose the architecture of domain adversarial learning fusing two pairwise learning classifiers. ‌During‌ adversarial training ‌of‌ sample features ‌from‌ the source and target domains‌, the classifiers also conducts adversarial learning. 
        \item We conduct rigorous cross-corpus tests using three published databases (SEED, SEED-IV, and SEED-V). All the proposed models achieved excellent performance. Among them, the McdPL achieved SOTA performance, while thorough analyses of model components and parameters, along with feature visualization, are performed to enhance our understanding of the results.
    \end{itemize}
    \indent The rest of this article is arranged as follows: We briefly described the background to the model in Sec.\ref{sec:related_work}. Then in Sec.\ref{sec:Methodology}, we introduce the concrete implementation of our proposed model in detail. Our experimental results was described in Sec.\ref{sec:experiment resules}. Finally, We discuss our model performance and conclusions in Sec.\ref{sec:Discussion_and_Conclusion}.
}
\end{CJK*}

\section{Related Work}
\label{sec:related_work}
\subsection{Domain Adversarial Learning}
    \indent Domain Adaptation Learning is a hot research field in machine learning, which maps samples in the source domain and target domain with different distributions into the same feature space so that their similar features in this space are as close as possible. It can effectively mitigate the learning problem of inconsistent probability distribution of source domain and target domain samples, so more and more researchers have‌ ‌paid attention to it, and it has been successfully applied in many fields \cite{Huang_Chen_Liu_Zheng_Tian_Jiang_2021,Zhu_Zhuang_Wang_Chen_Shi_Wu_He_2019,Gideon_McInnis_Provost_2021,Tzeng_Hoffman_Saenko_Darrell_2017,Luo_Zheng_Guan_Yu_Yang_2019,Gokhale_Anirudh_2022,lee2019sliced,Li_Liang_2023}.
    \\ \indent Such as, 
    Huang \etal \cite{Huang_Chen_Liu_Zheng_Tian_Jiang_2021} proposed a bi-hemisphere discrepancy convolutional neural network model (BiDCNN) for EEG emotion recognition, which can effectively learn the different response patterns between the left and right hemispheres.
    Zhu \etal \cite{Zhu_Zhuang_Wang_Chen_Shi_Wu_He_2019} presented a Multi-Representation Adaptation Network (MRAN), which dramatically improve the classification accuracy for cross-domain task and specially aims to align the distributions of multiple representations extracted.
    Gideon \etal \cite{Gideon_McInnis_Provost_2021} introduced Adversarial Discriminative Domain Generalization (ADDoG), which follows an easier to train 'meet in the middle' approach. This model iteratively moves representations learned for each dataset closer to one another, improving cross-dataset generalization.
    Tzeng \etal \cite{Tzeng_Hoffman_Saenko_Darrell_2017} presented an Adversarial Discriminative Domain Adaptation (ADDA) framework, which combines discriminative modeling, untied weight sharing and a GAN loss, achieving excellent unsupervised adaptation results on classification tasks.
    Luo \etal \cite{Luo_Zheng_Guan_Yu_Yang_2019} introduced a category-level adversarial network, aiming to enforce local semantic consistency during the trend of global alignment and align each class with an adaptive adversarial loss, improving semantic segmentation accuracy.
    Gokhale \etal \cite{Gokhale_Anirudh_2022} proposed an adversarial training approach which learns to generate new samples to maximize exposure of the classifier to the attribute-space, without having access to the data from the test domain,which enables deep neural networks to be robust against a wide range of naturally occurring perturbations.
    Lee \etal \cite{lee2019sliced} connected two distinct concepts for unsupervised domain adaptation: feature distribution alignment between domains by utilizing the task-specific decision boundary and the Wasserstein metric, which enhances the effectiveness and universality of the model.
    Zhou \etal \cite{Li_Liang_2023} proposed a transfer learning framework based on Prototypical Representation based Pairwise Learning (PR-PL) to encode semantic structures inherent in affective EEG data, aligning individual EEG features with a shared common feature space 
    However, PRPL only performs simple global alignment of sample features in domain anti-loss training, which may lose fine-grained information per class, which is more prominent in cross-corpus data. A visual example is shown in Fig.\ref{fig:pr_LMMD_cdd_methods}.(a), after global alignment, the distributions of the two domains are roughly the same, but the overly close feature distributions may degrade partial model performance‌.

\subsection{EEG-based emotion recognition}
    \indent The methods of emotion recognition are mainly based on non-physiological signals, such as facial expressions, language, Body gesture, and physiological signals, such as electroencephalogram (EEG), electrocardiogram (ECG), electromyogram (EMG), electrodermal activity (EDA), skin temperature (SKT), photoplethysmogram (PPG), respiration (RSP), electrooculogram (EOG) \cite{jerritta2011physiological,vora2019emotion,egger2019emotion,li2021physiological}. Compared to non-physiological signals, physiological signals seem to be more reliable in analyzing human emotions. As a kind of physiological signal, EEG has the advantage of being difficult to hide and disguise, so it has been widely studied in the field of affective computing, and the development of deep learning has greatly promoted research on emotion recognition based on EEG \cite{Song_Liu_Zheng_Zong_Cui_Li_Zhou_2023,ma2019emotion,Liu_Qiu_Zheng_Lu_2022,Anuragi_Singh_Sisodia_Bilas_Pachori_2022,Tao_Li_Song_Cheng_Liu_Wan_Chen_2023,Yin_Zheng_Hu_Zhang_Cui_2021,Li_Zheng_Zong_Cui_Zhang_Zhou_2021,Liu_Ding_Li_Cheng_Song_Wan_Chen_2020}. 
    \\ \indent Such as, 
    Song \etal \cite{Song_Liu_Zheng_Zong_Cui_Li_Zhou_2023} proposed a variational instance-adaptive graph method (V-IAG) that simultaneously captures the individual dependencies among different EEG electrodes and estimates the underlying uncertain information. 
    Ma \etal \cite{ma2019emotion} proposed a multimodal residual LSTM (MMResLSTM) network for emotion recognition. The MMResLSTM network shares the weights across the modalities in each LSTM layer to learn the correlation between EEG and other physiological signals, which can efficiently learn emotion-related high-level features. 
    Liu \etal \cite{Liu_Qiu_Zheng_Lu_2022} proposed two methods for extending the original DCCA model for multimodal fusion: weighted sum fusion and attention-based fusion, enhanceing recognition performance and robustness of emotion recognition models.
    Anuragi \etal \cite{Anuragi_Singh_Sisodia_Bilas_Pachori_2022} proposed an automated cross-subject emotion recognition framework based on EEG signals, which uses the Fourier-Bessel series expansion-based empirical wavelet transform (FBSE-EWT) method, and achieved excellent human emotion recognition performance.
    Tao \etal \cite{Tao_Li_Song_Cheng_Liu_Wan_Chen_2023} proposed an attention-based convolutional recurrent neural network (ACRNN) to extract more discriminative features from EEG signals and improve the accuracy of emotion recognition.
    Lin \etal \cite{Yin_Zheng_Hu_Zhang_Cui_2021} proposed a novel emotion recognition method based on a novel deep learning model (ERDL). The model fuses graph convolutional neural network (GCNN) and long-short term memories neural networks (LSTM), to extract graph domain features and temporal features.
    Liu \etal \cite{Liu_Ding_Li_Cheng_Song_Wan_Chen_2020} proposed an effective multi-level features guided capsule network (MLF-CapsNet), which incorporates multi-level feature maps learned by different layers in forming the primary capsules, for multi-channel EEG-based emotion recognition.

%------------------------------------------------------------------------------
\begin{figure*}[ht]
\centering
\subfloat{\includegraphics[width=1\textwidth]{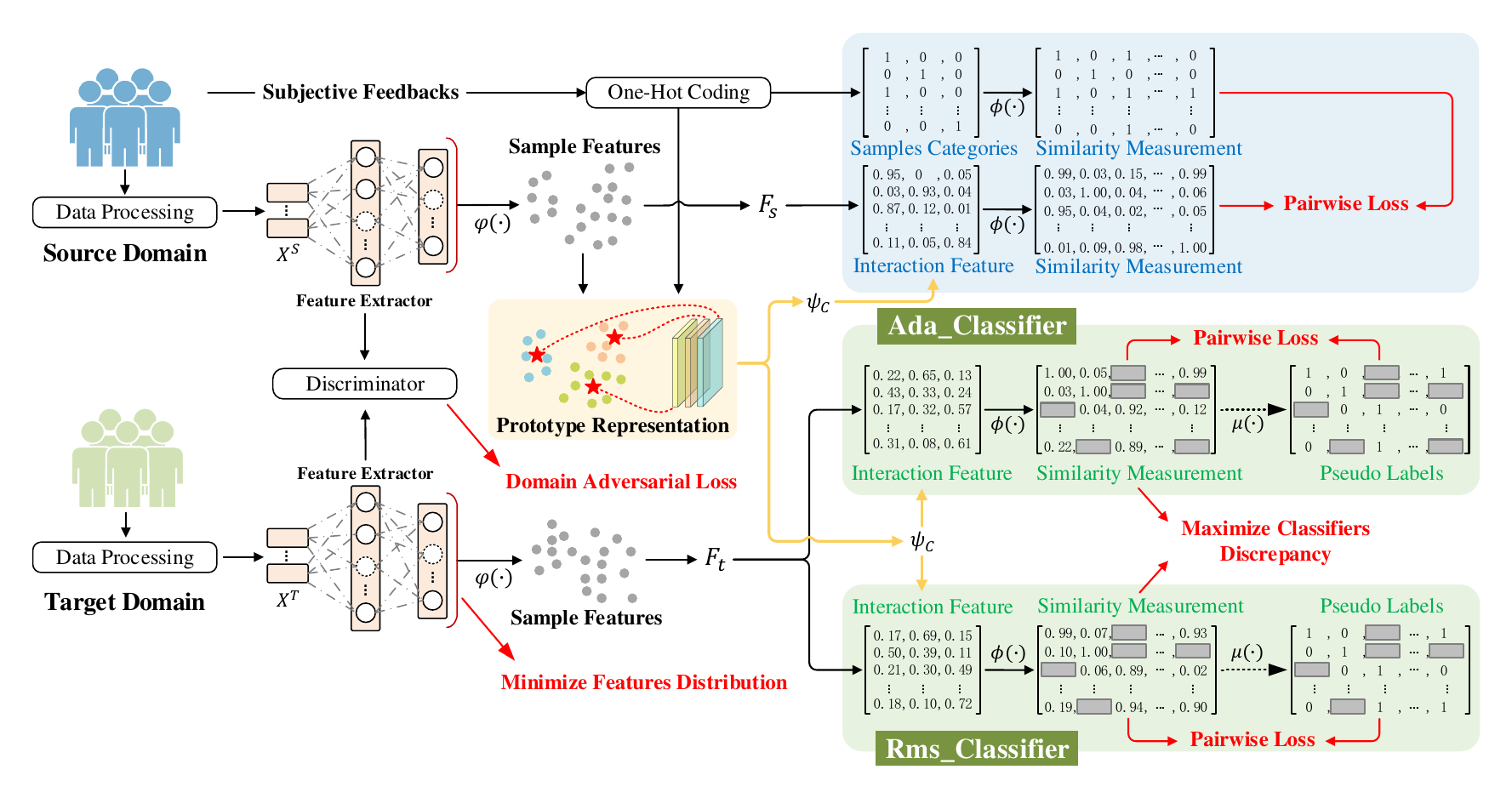}}
\caption{The McdPL model framework. The model includes Feature Discriminator, Prototype Representation Extraction and pairwise learning Classifiers (Ada\_classifier, Rms\_classifier). These modules include domain adversarial loss (Eq.\ref{Eq:1}), supervised and unsupervised pairwise learning losses for source and target domains (Eq.\ref{Eq:3}), maximum classifier discrepancy loss between Ada\_classifier and Rms\_classifier (Eq.\ref{Eq:18}), and minimum feature distribution loss (Eq.\ref{Eq:19}). $X^S$ and $X^T$ denotes source and target samples, respectively; $\varphi(\cdot)$ denotes the feature mapping operation; $F_s$ and $F_t$ represent the mapped features; $\psi_C$ denotes the prototype representation with emotion label \textit{c} (Eq.\ref{Eq:2}); $\phi(\cdot)$ denotes the similarity measurement between samples and $\mu(\cdot)$ denotes the pseudo-labeling operation.}
\label{fig:pr_MCD_methods}
\end{figure*}

\begin{table}[ht]
\centering
\caption{Frequently used notations and descriptions.}
\label{tab:T1}
\begin{tabular}{cc}
\bottomrule
\midrule
 Notation                    & Description           \\
\midrule
$\mathcal{S}\backslash\mathcal{T}$&Source\textbackslash{}Target Domain\\
$X^S\backslash X^T $ & Source\textbackslash{}Target Samples \\
$Y^S \backslash Y^T$ & Source\textbackslash{}Target True-Label  \\
$F_s \backslash F_t$ & Source\textbackslash{}Target Mapped Features  \\
$n \backslash m$     & Number of Source\textbackslash{}Target Samples\\
$D_f\left(\cdot\right)$  & Feature Distribution Discriminator \\ 
$D\left(\cdot\right)$    & Domain Discriminator \\ 
$R\left(\cdot\right)$    & Rounding Function\\
$f\left(\cdot\right)$    & Feature Extractor  \\
$\phi\left(\cdot\right)$ & Similarity Measurement\\
$|\cdot| $               & Absolute-Value Operation\\
$\mathcal{H\kappa}$      & \ \ \ \ \ \ \ \ Reproducing Kernel Hillbert Space\ \ \ \ \ \ \ \ \\
$\underline{C}$          & Number of Emotion Labels\\
$\Gamma$                 & Interaction Feature  \\
$\psi$                   & Prototype Representation  \\
$\mu$                    & Pseudo-Labels Computation\\
$\varphi $               & Samples Feature Mapping\\
\midrule
\bottomrule
\end{tabular}
\end{table}
%------------------------------------------------------------------------------
\section{Methodology}
\label{sec:Methodology}
\begin{CJK*}{UTF8}{gbsn}{
    \indent Suppose the samples and labels of source domain $\mathcal{S}$ and target domain $\mathcal{T}$ are expression by $(X^S,Y^S)$ and $(X^T,Y^T)$, and $(X^S,Y^S)=(x^s_i,y^s_i)^n_{i=1}$, $(X^T,Y^T)=(x^t_i,y^t_i)^m_{i=1}$, where $x^s_i$ and $x^t_i$ are EEG data samples, $y^s_i$ and $y^t_i$ are corresponding emotional labels. $n$ and $m$ are the number of samples in the source domain and the target domain, respectively. Significantly, the label information of the target domain is completely missing in the training process. For clearer expression, frequently used notations in this paper are summarized in Tab.\ref{tab:T1}.

\subsection{Underlying Architecture}
    \indent Our proposed model consists of three main modules: feature discriminator method, prototype  representation extraction method and pairwise learning classifier. 
\subsubsection{Feature Discriminator}
    \indent Based on domain adaptive neural network (DANN), We introduce domain adversarial training to characterize the sample features. This method utilizes domain adaptive and adversarial training to minimize the differences between sample feature representations and effectively extract domain invariant attributes \cite{Li_2021_eeg,Li_2022_eeg,Du_2022_eeg,pi_2018_uns,Li_Zheng_2022_eeg}. Thus, To distinguish whether the sample features belong to the source domain or the target domain, suppose that $D(\cdot)$ represents the domain discriminator and $f(\cdot)$ represents the feature extractor. The domain countermeasure loss function $l_{disc}$ adopts the binary cross-entropy loss function, which is defined as:
    \begin{equation}
    \label{Eq:1}
    \begin{split}
        l_{disc}(X^S,X^T) = -\sum_{i=0}^{n}  \log{\left [ D(f(x_i^s)) \right ] }
        -\sum_{j=0}^{m}\log{\left [1- D(f(x_j^t))  \right ] }
    \end{split}
    \end{equation}
    here,  $x^s_i$ and $x^t_j$ represent the $i^{th}$ sample from the source domain and the $j_{th}$ sample from the target domain, respectively. The $n \backslash m$ denote the number of samples in the source and target domains, respectively. 

\subsubsection{Prototype Representation Extraction}
    \indent For each emotion, it is assumed that there is a prototype representation, which represents the essential attribute of the emotion category, and samples belonging to the same category are distributed around the prototype representation. From the perspective of probability distribution, the prototype representation can be regarded as the ‌centroid‌ of the features of all samples within that category. Therefore, extracting prototype representations can enhance the generalization ability of the model. Suppose that $X^S_c=\{x^{sc}_i\}^{[X^S_c]}_{i=1}$, where $x_i^{sc}$ denotes the $i_{th}$ sample belonging to category $c$ in the source domain, and \([X_c^S]\) represents the number of samples in that category. $\underline{C}$ represent the number of emotional categories. For convenience, the prototype representations for each emotion can be defined as $\{\psi_1,\psi_2,\dots,\psi_{\underline{C}}\}=\psi_{1:\underline{C}}$. Thus, the prototype representation defined as :
    \begin{equation}
    \label{Eq:2}
    \psi_{1:\underline{C}} =  \frac{1}{\left [ X_{c}^{S} \right ]}\sum_{x_i^{sc}\in X_{c}^S}^{}f(x_{i}^{sc}),
    \end{equation}
    overall, the prototype representation for the emotion category $c$ is obtained by the average vector of all sample features belonging to that category. During model training, the prototype representations are constantly updated iteratively, ultimately yielding the optimal prototype representation for each emotion category.
    
\subsubsection{Pairwise Learning Classifier}
\label{sec:Pairwise Learning}
    \indent Traditional point-to-point learning only focuses on the relationship between a single sample feature ($F$) and a prototype representation ($\psi$), but in order to capture the internal relationship between multiple samples, we adopt a pairwise learning strategy. Assume that the interaction feature $\Gamma$ represents the interaction between the sample feature and the prototype representation, defined as $\Gamma=(F\cdot\psi_c\cdot \theta)$, where (·) represents the inner product operation, $F$ represents the extracted sample feature, $\psi_c$ is the prototype representation with emotion class c, $\theta$ is a trainable transformation matrix. Therefore, the pairwise learning target loss function based on interaction features is defined as:
    \begin{equation}
    \label{Eq:3}
    \begin{split}
       l_{pair } = \frac{1}{(n\backslash m)^2} \sum_{i,j\in{k}}^{} \left[-\mu_{ij} \log{ \phi_{ij}  (\Gamma)  }-(1-\mu_{ij})\log{\left (1- \ \phi_{ij}  (\Gamma  ) \right )}\right] 
    \end{split}
    \end{equation}
    here, $\phi_{ij}(\cdot)$ represents the similarity measurement between samples $x_i$ and  $x_j$, indicating the probability that samples $x_i$ and  $x_j$ belong to the same emotion label, ranges from 0 to 1, with values closer to 1 suggesting that samples $x_i$ and  $x_j$ are more likely to belong to the same emotional category, and vice versa. $\mu_{ij}$ indicates whether samples $x_i$ and  $x_j$ belong to the same emotion category, with the value 0 or 1. $\mu_{ij} = 1$ indicates that the sample pair belongs to the same emotion category, and vice versa. In the unsupervised learning of the target domain, $\mu_{ij}$ is determined by pseudo-label. We set two thresholds, if the similarity measurement is higher than the upper threshold, the sample pairs are considered to belong to the same category, while if the similarity measurement is lower than the lower threshold, the sample pairs are not considered to belong to the same category.
    
\subsection{Local maximum mean discrepancy with Pairwise Learning}
\label{sec:lmmd}
    \indent Maximum mean discrepancy (Mmd) is a statistical measure used to quantify the difference between two distributions. In transfer learning, Mmd can be applied to evaluate the similarity between the source domain and the target domain samples extracted by the feature extractor, that is the non-parametric measure \cite{Gretton_Borgwardt_2012}. When applied to the loss function, the optimization objective is to minimize this metric. Mmd maps the feature distribution from a low-dimensional space to a high-dimensional space. The samples evaluate their similarity by calculate the distance between their distributions, thereby enabling fine-grained alignment of source and target domain samples in the feature space.The source and target domain samples are represented as $X^S=\{x^s_i\}^{n}_{i=1}$ and $X^T=\{x^t_j\}^{m}_{j=1}$, respectively. The objective function for Mmd is defined as:
    \begin{equation}
    \label{Eq:9}
    \begin{split}
        {MMD_{\mathcal{H\kappa}}}(X^{S}, X^{T}) = \left\| \frac{1}{n} \sum_{i=1}^n \varphi(x_i^{s}) - \frac{1}{m} \sum_{j=1}^m \varphi(x_j^{t}) \right\|_{\mathcal{H\kappa}}^2
    \end{split}
    \end{equation}
    here, $\varphi(\cdot)$ denotes the feature mapping operation for the source domain sample $x^s_i$ or the target domain sample $x^t_j$. $\mathcal{H\kappa}$ represents the corresponding Reproducing Kernel Hilbert Space (RKHS) with feature kernel $\kappa$, which is a feature mapping space. The feature kernel $\kappa$ can be understood as a distance function, which defines the distance between samples $x^s$ and $x^t$ in the RKHS. This distance measure can be obtained by calculating the inner product of the samples in the feature space, expressed as $\kappa=\left \langle {\varphi(x^{s}),\varphi(x^{t})} \right \rangle $. 
    \\ \indent Furthermore, as shown in Fig.\ref{fig:pr_LMMD_cdd_methods}.(b), we focus not only on macro-level alignment but also on aligning sample features from a micro perspective, and needs to consider the relationships between subdomains of the same class across different domains. Therefore, based on Mmd, we propose the LmmdPL, suppose that the probability weights be denoted as $\omega$, the Lmmd is defined as :
    \begin{equation}
    \label{Eq:10}
    \begin{split}
        {Lmmd_{\mathcal{H\kappa}}}(X^{S}, X^{T}) =\frac{1}{\underline{C}} \sum_{c=1}^{\underline{C}}{  \left\|  \sum_{i=1}^{n^c} {\omega _i^{sc}\varphi(x_i^{s})} -  \sum_{j=1}^{m^c} {\omega _j^{tc}\varphi(x_j^{t})} \right\|_{\mathcal{H\kappa}}^2}
    \end{split}
    \end{equation}
    here, $n^c$ and $m^c$ represent the number of samples belonging to category $c$ in the source and target domains, respectively. $\omega _i^{sc}$ and $\omega _j^{tc}$ indicate the probabilities of the source domain sample $x^s_i$ and the target domain sample $x^t_j$ belonging to their respective domains given category $c$. The Lmmd helps balance the characteristic contributions of the two domains, and focusing on the relationship between subdomains of the same class across different domains. Significantly, LmmdPL also requires mapping the sample features into the RKHS with the feature kernel $\kappa$. Therefore, Eq.\ref{Eq:10} can be transformed into:
    \begin{equation}
    \label{Eq:11}
    \begin{split}
        Lmmd_{\mathcal{H\kappa}}(X^{S}, X^{T})=\frac{1}{\underline{C}}\sum_{c=1}^{\underline{C}} \left [] \sum_{i=1}^{n}\sum_{j=1}^{n} \omega_i^{sc}\omega_j^{sc}\kappa(x_i^s,x_j^s)\right.
        \\ \left.+\sum_{i=1}^{m}\sum_{j=1}^{m} \omega_i^{tc}\omega_j^{tc}\kappa(x_i^t,x_j^t)-2\sum_{i=1}^{n}\sum_{j=1}^{m} \omega_i^{sc}\omega_j^{tc}\kappa(x_i^s,x_j^t)\right ] 
    \end{split}
    \end{equation}
    \indent In the source domain samples, we can easily obtain the true labels, But in the target domain, the label information is completely missing in the training process. Thus, we use the interaction feature $\Gamma$ to approximate the estimation of the emotional categories of the samples, as $\Gamma$ effectively represents the probability of the target domain samples belonging to each emotional category. Consequently, it is appropriate to calculate the weights $\omega$ using the interaction feature $\Gamma$ in the target domain.
    Significantly, the sum of the weights of all samples belonging to category \( c \) in  the source or target domain is always equal to 1, $\sum_{i=1}^{n^c}\omega _i^{sc} = 1$ and $\sum_{j=1}^{m^c}\omega _j^{tc} = 1$. Thus, the weights $\omega _i^{sc}$ and $\omega _j^{tc}$ for the source and target domain samples are defined as follows:
    \begin{equation}
    \label{Eq:12}
        \omega _i^{sc} = \frac{y_i^c}{\sum_{k=1}^{n^c}y_k^c} , \omega _j^{tc} = \frac{\Gamma _j^c}{\sum_{k=1}^{m^c}\Gamma _k^c}
    \end{equation}
    here, $y^c_i$ represents the entry at the $c_{th}$ position of the true label  $y_i$ for the source domain sample $x_i^s$. Significantly, the true labels for samples in the source domain are encoded using one-hot encoding. Eq.\ref{Eq:11} is defined as the loss function of lmmd ($l_{lmmd\_\mathcal{H\kappa}}$), Eq.\ref{Eq:3} is defined as the pairwise learning loss function of supervised source domain and unsupervised target domain respectively ($l_{pair}^s \backslash l_{pair}^t$), and Eq.\ref{Eq:1} is defined as the domain adversarial loss function ($l_{disc}$). Overall, the objective function of LmmdPL is defined as:
    \begin{equation}
    \label{Eq:13}
        L_{Lmmd}= l_{pair}^s+\alpha l_{pair}^t-\beta l_{disc} + \gamma  l_{lmmd\_\mathcal{H\kappa}}
    \end{equation}
    
\subsection{Contrastive domain discrepancy with Pairwise Learning}
\label{sec:cdd}
    \indent LmmdPL method achieves alignment of subdomains in the feature space by reducing the distance between subdomains of the same category across the two domains. However, the distances between feature distributions of different categories in different domains may be too close. In the source domain, guided by true labels, the classifier can calculate complex decision boundaries to fit samples of different categories. However, when this rule is transferred to the target domain, samples located near the boundary are maybe misclassified by the model.
    \\ \indent Thus, we further propose Contrastive Domain Discrepancy. Suppose $d^{cc}(\cdot)$ represent the intra-domain discrepancy among samples of the same category, and $d^{cc'}(\cdot)$ represent the inter-domain discrepancy between different categories. For convenience, the predicted labels of the target domain samples $\{p^t_1,p^t_2,\dots,p^t_m\}\in \Gamma^t $ are expressed as $p^t_{1:m}$. Overall, the Cdd objective function is defined as：
    \begin{equation}
    \label{Eq:14}
    \begin{split}
        {CDD_{\mathcal{H\kappa}}} = \frac{1}{\underline{C}} \sum_{c=1}^{\underline{C}} d^{cc}(p^t_{1:m}, \varphi)- \frac{1}{\underline{C}(\underline{C}-1)} \sum_{c=1}^{\underline{C}} \sum_{\substack{c' = 1 \\ c' \neq c}}^{\underline{C}} d^{cc'}(p^t_{1:m}, \varphi)
    \end{split}
    \end{equation}
    here, $\varphi$ denotes the feature mapping operation. Significantly, the objective function consists of two parts, the portion before the minus sign indicates the inter-domain discrepancy among samples of the same category, while the latter represents the inter-domain discrepancy among samples of different categories. In this strategy, intra-class and inter-class discrepancies are optimized in opposing directions. With the iterative improvement of the model, as shown in Fig.\ref{fig:pr_LMMD_cdd_methods}.(c), the feature representation of samples of the same emotion category becomes more compact, while those of different emotional categories gradually move away from each other. which can enhance the processing ability of the model for samples near the decision boundary. Suppose $\gamma$ is a hyperparameter. We take  Eq.\ref{Eq:14} is defined as the loss function of Cdd ($l_{cdd\_\mathcal{H\kappa}}$).
    Thus, the objective loss function of CddPL defined as:
    \begin{equation}
    \label{Eq:15}
        L_{Cdd}= l_{pair}^s+\alpha l_{pair}^t-\beta l_{disc} + \gamma  l_{cdd\_\mathcal{H\kappa}}
    \end{equation}

\subsection{Maximum classifier discrepancy with Pairwise Learning}
\label{sec:mcd}
%----------------------------------------------------------------------------
\begin{figure}
\centering
\subfloat{\includegraphics[width=0.5\textwidth]{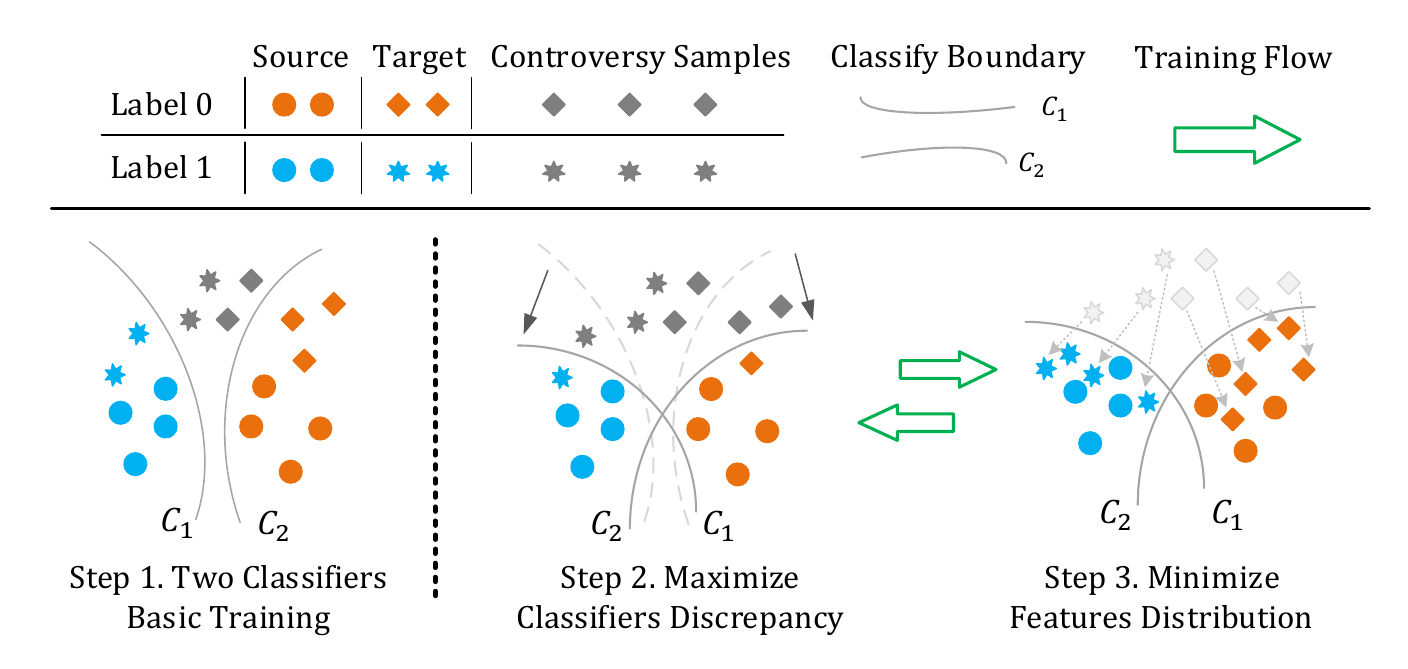}}
\caption{McdPL model training steps. The gray samples represent the controversial samples, which usually achieves different results in the two classifiers. The primary objective of Step 2 is to identify more controversial samples. Step 3 focuses on adjusting the mapping of controversial samples in the feature space to minimize feature distribution discrepancy.}
\label{fig:pr_MCD_method_II}
\end{figure}

\begin{figure}
\centering
\subfloat{\includegraphics[width=0.5\textwidth]{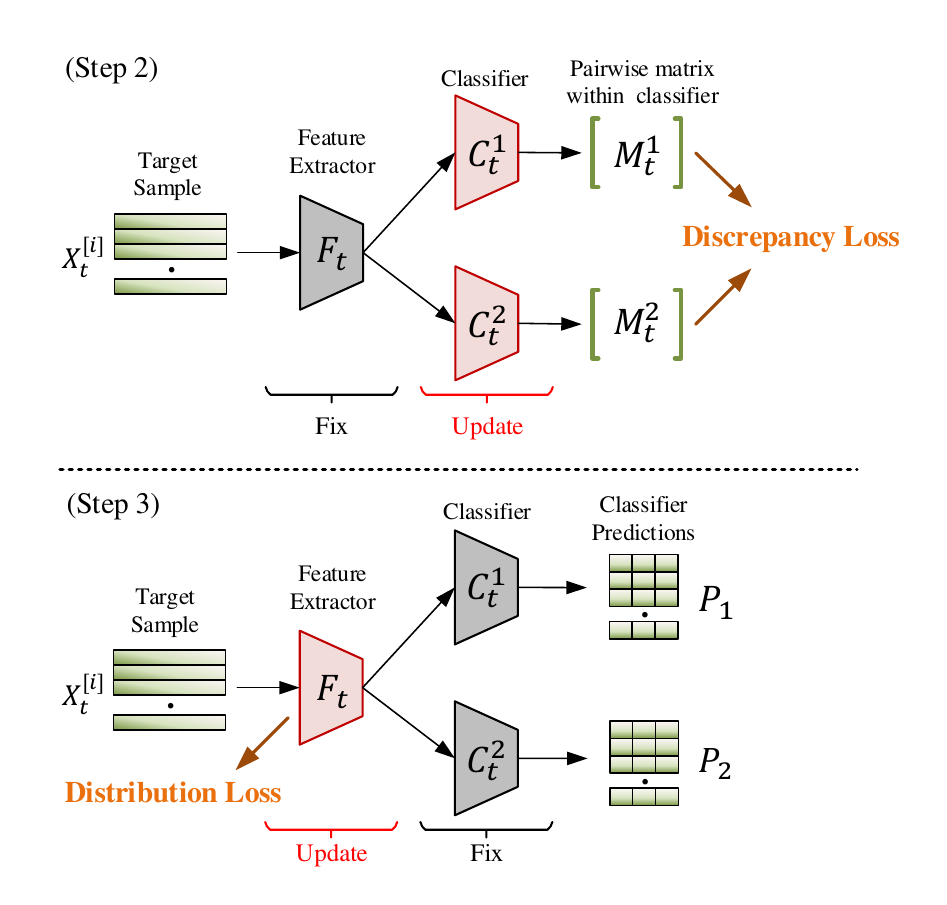}}
\caption{\textcolor{black}{Schematic diagram of each module of McdPL model training. Step 2 corresponds to Fig.\ref{fig:pr_MCD_method_II}.(Step 2), in which the parameters of the feature extractor are fixed and two classifiers are trained. Step 3 corresponds to Fig.\ref{fig:pr_MCD_method_II}.(Step 3), in which the parameters of the two classifiers are fixed and  the parameters of the feature extractor are trained.}}
\label{fig:pr_MCD_method_III}
\end{figure}
%----------------------------------------------------------------------------

%-------------------------------------------------------------------------McdPL_algorithm
\begin{algorithm}[ht]
\caption{McdPL model Training algorithm.}
\label{fig:mcd_algorithm}
\textbf{Input:} Source and Target domain samples $X^S\backslash X^T $;
Source and Target Labels $Y^S \backslash Y^T$;
Feature Extractor $f_s(\cdot)\backslash f_t(\cdot)$;
Prototype Representation Extractor $\psi_c$;
Interaction Features $\Gamma_{Ada}\backslash \Gamma_{Rms}$;
domain adversarial loss $\mathcal{L}_{dise}$; 
Source and Target domain pairwise loss $\mathcal{L}_{pair}^s\backslash \mathcal{L}_{pair}^t$; 
Discrepancy between classifiers $\mathcal{D}$;
Discrepancy of Samples Feature $\mathcal{D}_f$;
Hyperparameter $M$;
Accuracy and Standard-Deviation $Acc \backslash Std$.\\
\textbf{Training:}
%\textbf{for} $n = 1$ \textbf{to} $(Epoch\_size)$ \textbf{do}:
\begin{algorithmic}[1]
    \For{$n = 1$ \textbf{to} $(Epoch\_size)$}:
        \State \textbf{Step 1}
        %\State \textcolor{black}{\#Domain Adversarial Loss}
        \State $l_{disc}=\mathcal{L}_{disc}(f(X^S),f(X^T))$\textcolor{gray}{\#Adversarial Loss}
        %\State \textcolor{black}{\#Source Pairwise in two Classifiers}
        \State $l_{pair}^s = \mathcal{L}_{pair}^s(f(X^S),Y^S)$\textcolor{gray}{\#Source Pairwise Loss}
        \State \textcolor{gray}{\#Target Pairwise Loss}
        \State $l^t_{pair\_ada},l^t_{pair\_rms} = \mathcal{L}^t_{pair}(F_t,\psi_c,\theta )$
        \State $L_1 = l_{disc}+\alpha l_{pair}^s+\beta l^t_{pair\_ada}+\gamma l^t_{pair\_rms}$
        \State \textbf{Step 2}
        \State \textcolor{black}{Lock $f_s(X^S)\backslash f_t(X^T)$ }
        \State $l^t_{pair\_ada},l^t_{pair\_rms} = \mathcal{L}^t_{pair}(F_t,\psi_c,\theta )$
        %\State \textcolor{black}{\#Classifiers Discrepancy}
        \State $d_{cla} = \mathcal{D}(\Gamma_{Ada},\Gamma_{Rms})$\textcolor{gray}{\#Classifiers Discrepancy}
        \State $L_2 = l^t_{pair\_ada}+l^t_{pair\_rms}-d_{cla}$
        \State \textcolor{black}{UnLock $f_s(X^S)\backslash f_t(X^T)$}
        \State \textbf{Step 3}
        \State \textcolor{black}{Lock $Ada\_classifier\backslash Rms\_classifier$}
        \For{$m = 1$ \textbf{to} $M$}:
            \State $l_{disc}=\mathcal{L}_{disc}(f(X^S),f(X^T))$\textcolor{gray}{\#Adversarial Loss}
            \State $l_{pair}^s = \mathcal{L}_{pair}^s(f(X^S),Y^S)$\textcolor{gray}{\#Source Pairwise Loss}
            \State $L_3 = l_{disc} + l_{pair}^s$\textcolor{gray}{\#Features Distribution}
        \EndFor
        \State UnLock $Ada\_Classifier\backslash Rms\_Classifier$
    \EndFor
\end{algorithmic}
%\textbf{end for}\\
\textbf{Evaluation:} Acc\textbackslash Std = Predicts($X^T$,$Y^T$)
\end{algorithm}
%---------------------------------------------------------------------------------------
    \indent Both LmmdPL and CddPL models are built from the perspective of domain adaptation. Furthermore, from the perspective of adaptive optimization of rule domain, in order to find a finer grained matching method between source domain and target domain, we consider the relationship between classification boundary and target samples. Thus, we designed two classifiers, named the Ada classifier and the Rms classifier, as shown in Fig.\ref{fig:pr_MCD_methods}. The Ada classifier employs the Adaptive Moment Estimation (Adam\cite{Kingma_Ba_2014}) gradient descent algorithm, while the Rms classifier utilizes the Root Mean Square Propagation (RMSprop\cite{Ruder_2016}) gradient descent algorithm.
    Although the classifier structures are identical, they obtain different parameters from the outset of training. During training, when the sample features from the source domain and the target domain are subjected to domain adversarial learning, the two classifiers also form adversarial learning.
    \\ \indent In the feature space, suppose that there is an implicit correlation between the distribution of target domain samples and the corresponding distribution of source domain. Specifically, when the classification boundary derived from the source domain is directly applied to the target domain, it may yield classification results by both high and low confidence. The target domain samples with high confidence are closer to the source domain samples in feature space, indicating target domain exist source domain support. In contrast, samples with lower confidence are typically positioned further away from all feature cluster centers and are often located near the decision boundary, we refer to these samples as controversial samples. Identifying these controversial samples is important to improve the performance of the model for emotion recognition across databases, and these samples usually produce different classification results between the two classifiers. Thus, we propose a method to maximize classification differences during training. Without this step, the two classifiers would become excessively similar. Subsequently, the feature extractor is employed to deceive the discriminator, ensuring that target domain samples are generated within the support range of the source domain, thereby minimizing the feature distribution discrepancy. Importantly, the model iterates through this adversarial learning step repeatedly. The McdPL related algorithms are presented in Alg.\ref{fig:mcd_algorithm}. Overall, the model training is organized into the following three steps.

    \subsubsection{Step 1: Basic Model Training}
    \label{sec:bt}
        \indent As shown in Fig.\ref{fig:pr_MCD_method_II}.(Step 1), In order to enable the model to learn the basic classification parameters, we conduct preliminary training on the model's feature generator $(F)$ and classifier $(C_{Ada},C_{Rms})$, aiming to obtain the baseline performance of the model in emotion recognition. The step 1 loss function is defined as:
        \begin{equation}
        \label{Eq:16}
            L_{1} = l_{pair}^s+\alpha  l_{pair\_ada}^t+\beta  l_{pair\_ams}^t+\gamma  l_{dise}
        \end{equation}
        here, $l_{pair\_ada}^t$ and $l_{pair\_ams}^t$ represent the pairwise learning  losses of the Ada\_classifier and Rms\_classifier, respectively. $\alpha,\beta,\gamma$ are hyperparameters optimized for model performance.

    \subsubsection{Step 2: Maximize Classifiers Discrepancy}
    \label{sec:Mcd}
        \indent As shown in Fig.\ref{fig:pr_MCD_method_II}.(Step 2) and Fig.\ref{fig:pr_MCD_method_III}.(Step 2), the model freezes the parameters of the feature generator $(F)$ and focuses on training the two classifiers $(C_{Ada}\backslash C_{Rms})$ during training. This step aims to maximize the discrepancy between the classifiers, ensuring accurate classification of source domain samples while simultaneously identifying a greater number of target domain controversial samples. Therefore, we introduce a metric that quantifies the discrepancy between the two classifiers, defined as:
        \begin{equation}
        \label{Eq:17}
            l_{diff}(X^T)=Mean \left [   \sum_{x^t\in X^T}^{}\left | \phi (\Gamma _{Ada}(x^t))-\phi (\Gamma _{Rms}(x^t)) \right |\right ]
        \end{equation}
        here, $x^t\in X^T$ denotes a target domain samples, $|\cdot|$ represents the absolute operation. $\Gamma_{Ada}(x^t)$ and $\Gamma_{Rms}(x^t)$ denote the predicted probabilities for sample  $x^t$  by the Ada classifier and Rms classifier, respectively. $\phi(\cdot)$ is a similarity measurement between samples pairs. 
        Suppose that $l_{diff}$  represents the discrepancy value, the loss function for step 2 is defined as :
        \begin{equation}
        \label{Eq:18}
            L_{2} = \min(L_{1}-max_{\{C_{Ada},C_{Rms}\}}l_{diff}(X^T))
        \end{equation}
        here, during the model training process, the overall trend of this loss function is downward. Consequently, $L_1$ and $l_{diff}(X^T)$ are optimized in opposite directions. It is worth noting that $L_2$ remains greater than 0.

    \subsubsection{Step 3: Minimize Features Distribution}
    \label{sec:mfd}
        \indent As shown in  Fig.\ref{fig:pr_MCD_method_II}.(Step 3) and Fig.\ref{fig:pr_MCD_method_III}.(Step 3), in the process of model training, the goal is to minimize the difference of feature distribution. Therefor, we freezes the parameters of the two classifiers $(C_{Ada} \backslash C_{Rms})$, and training  the feature generator $(F)$ to minimize the feature distribution. The model adjusts the mapping of target domain samples during training, the features extracted from the target domain samples gradually converge towards the clustering centers of the source domain in the feature space, enabling the Ada classifier and Rms classifier to achieve more accurate classifications. The  loss function for step 3 is defined as:
        \begin{equation}
        \label{Eq:19}
            L_3 = \min_{F}\left (l_{disc} + l_{pair}^s  \right ) 
        \end{equation}
        here, $l_{disc}$ represents the domain discriminator loss of the model, and $l_{pair}$ represents the pairwise learning loss of the source domain. Minimizing the feature distribution is significant for the controversy samples identified in step 2. 
}\end{CJK*}

%--------------------------------------------------------------------------表格的结果
\begin{table*}[ht]
\begin{center}
\caption{The accuracy and Standard-Deviations of the various EEG emotion recognition Methods in Cross-corpus Cross-subjects Single-session hold-out-valuation, the results are shows as (Accuracy\% $\pm$ Standard-Deviation\%). The database training combination is represented as (Source Domain $\to$ Target Domain). \textcolor{orange}{$\uparrow\%$} denotes the improvement in accuracy between the best performance and the second-best performance. Here, the model results reproduced by us are indicated by '*'.}
\label{tab:test_session_1}
\scalebox{0.80}{
\color{black}
\begin{tabular}{lccccccc}
\bottomrule
\midrule
Methods & SEED$\to$SEED-IV                  & SEED$\to$SEED-V 
        & SEED-IV$\to$SEED                  & SEED-IV$\to$SEED-V
        & SEED-V$\to$SEED                   & SEED-V$\to$SEED-IV
        & Ave\_Acc\\
\midrule
SVM*\cite{SVM}     
        &  43.14 $ \pm $ 8.36               &  39.08 $ \pm $ 9.08  
        &  44.77 $ \pm $ 8.62               &  31.44 $ \pm $ 8.56  
        &  37.27 $ \pm $ 7.00               &  35.97 $ \pm $ 7.27  
        &  38.45\\
K-Means*\cite{K_Means} 
        &  48.57 $ \pm $ 8.62               &  45.33 $ \pm $ 8.62  
        &  50.87 $ \pm $ 6.35               &  43.37 $ \pm $ 7.74  
        &  42.70 $ \pm $ 6.65               &  46.21 $ \pm $ 7.60  
        &  46.18\\
DANN*\cite{DANN}    
        &  44.13 $ \pm $ 7.85               &  44.92 $ \pm $ 8.85  
        &  51.07 $ \pm $ 8.25               &  27.78 $ \pm $ 8.52  
        &  41.53 $ \pm $ 7.13               &  29.93 $ \pm $ 7.17  
        &  39.89\\
KNN*\cite{KNN}    
        &  41.35 $ \pm $ 4.63               &  36.34 $ \pm $ 4.52  
        &  35.09 $ \pm $ 4.30               &  26.02 $ \pm $ 4.44  
        &  30.63 $ \pm $ 4.34               &  35.63 $ \pm $ 4.45  
        &  34.18\\
LeNet*\cite{LeNet}    
        &  42.30 $ \pm $ 8.55               &  35.60 $ \pm $ 8.63  
        &  51.61 $ \pm $ 8.11               &  33.51 $ \pm $ 7.08  
        &  41.26 $ \pm $ 7.15               &  35.92 $ \pm $ 8.03  
        &  40.03\\
LmmdPL  
        &  51.04 $ \pm $ 2.82               &   \underline{54.71} $ \pm $ 3.34  
        &  \underline{59.15} $ \pm $ 0.96   &  49.93 $ \pm $ 2.93  
        &  \underline{58.27} $ \pm $ 4.57               &  49.94 $ \pm $ 4.30  
        &  53.76\\
DCORAL*\cite{DCORAL}    
        &  41.48 $ \pm $ 8.50               &  46.65 $ \pm $ 9.20  
        &  54.12 $ \pm $ 8.36               &  33.99 $ \pm $ 8.20  
        &  44.10 $ \pm $ 6.80               &  32.30 $ \pm $ 8.38  
        &  42.12\\
DDA*\cite{DDA}    
        &  48.28 $ \pm $ 8.31               &  46.01 $ \pm $ 9.33  
        &  56.26 $ \pm $ 4.96               &  47.29 $ \pm $ 9.11  
        &  44.26 $ \pm $ 4.97               &  41.22 $ \pm $ 7.70  
        &  47.22\\
PrPL*\cite{Li_Liang_2023}    
        &  51.08 $ \pm $ 1.08               &  52.71 $ \pm $ 3.34  
        &  58.53 $ \pm $ 1.91               &  47.58 $ \pm $ 2.95  
        &  55.54 $ \pm $ 4.20               &  50.87 $ \pm $ 2.71  
        &  52.72\\
CddPL   
        &  \underline{52.63} $ \pm $ 1.59   &  53.92 $ \pm $ 2.19  
        &  58.53 $ \pm $ 1.96               &  \underline{51.74} $ \pm $ 3.51  
        &  56.28 $ \pm $ 2.92   &  \underline{51.93} $ \pm $ 3.11  
        &  \underline{54.17}\\
BLFBA*\cite{BLFBA}    
        &  46.24 $ \pm $ 9.03               &  40.95 $ \pm $ 8.90  
        &  52.17 $ \pm $ 6.82               &  34.79 $ \pm $ 8.73  
        &  44.52 $ \pm $ 8.12               &  37.12 $ \pm $ 8.04  
        &  42.63\\
McdPL   
        &  \textbf{53.50 $ \pm $ 2.48}      &  \textbf{55.68 $ \pm $ 1.69}  
        &  \textbf{67.92 $ \pm $ 1.03}      &  \textbf{53.36 $ \pm $ 5.59}
        &  \textbf{68.75 $ \pm $ 9.35}      &  \textbf{54.38 $ \pm $ 3.70}  
        &  \textbf{58.93}\\
\midrule 
        & \textcolor{orange}{$\uparrow$ 1.13} 
        & \textcolor{orange}{$\uparrow$ 0.97}
        & \textcolor{orange}{$\uparrow$ 8.77} 
        & \textcolor{orange}{$\uparrow$ 1.62}
        & \textcolor{orange}{$\uparrow$ 10.48} 
        & \textcolor{orange}{$\uparrow$ 2.45}
        & \textcolor{orange}{$\uparrow$ 4.76}\\
%\multicolumn{2}{l}{\textbf{PADDA}} & \textbf{93.06/05.12}  \\
\midrule
\bottomrule
\end{tabular}
}
\end{center}
\end{table*}

\begin{table*}[ht]
\begin{center}
\caption{The accuracy and Standard-Deviations of the various EEG emotion recognition Methods in Cross-corpus Cross-subjects Cross-session hold-out-valuation, the results are shows as (Accuracy\% $\pm$ Standard-Deviation\%). The database training combination is represented as (Source Domain $\to$ Target Domain). \textcolor{orange}{$\uparrow\%$} denotes the improvement in accuracy between the best performance and the second-best performance. Here, the model results reproduced by us are indicated by '*'.}
\label{tab:test_all_session}%\underline{}
\scalebox{0.8}{
\color{black}
\begin{tabular}{lccccccc}
\bottomrule
\midrule
Methods & SEED$\to$SEED-IV                  & SEED$\to$SEED-V 
        & SEED-IV$\to$SEED                  & SEED-IV$\to$SEED-V
        & SEED-V$\to$SEED                   & SEED-V$\to$SEED-IV
        & Ave\_Acc\\  
\midrule
SVM*\cite{SVM}     
        &  42.96 $ \pm $ 8.82               &  40.07 $ \pm $ 8.07  
        &  41.23 $ \pm $ 9.18               &  38.21 $ \pm $ 9.24  
        &  42.82 $ \pm $ 8.17               &  38.08 $ \pm $ 7.39  
        &  40.66\\
K-Means*\cite{K_Means} 
        &  47.92 $ \pm $ 8.02               &  44.21 $ \pm $ 8.87  
        &  51.01 $ \pm $ 6.91               &  44.42 $ \pm $ 8.06  
        &  45.77 $ \pm $ 6.87               &  46.95 $ \pm $ 8.00  
        &  46.71\\
DANN*\cite{DANN}    
        &  45.60 $ \pm $ 8.30               &  48.04 $ \pm $ 8.21  
        &  51.18 $ \pm $ 8.05               &  39.50 $ \pm $ 8.24  
        &  49.60 $ \pm $ 9.08               &  46.89 $ \pm $ 7.68  
        &  46.80\\
KNN*\cite{KNN}    
        &  41.94 $ \pm $ 4.65               &  43.24 $ \pm $ 4.41  
        &  40.92 $ \pm $ 4.28               &  38.99 $ \pm $ 4.39  
        &  34.92 $ \pm $ 4.33               &  36.15 $ \pm $ 4.43  
        &  39.36\\
LeNet*\cite{LeNet}    
        &  46.49 $ \pm $ 8.58               &  39.75 $ \pm $ 6.55  
        &  47.21 $ \pm $ 8.68               &  38.87 $ \pm $ 7.82  
        &  46.85 $ \pm $ 8.44               &  41.27 $ \pm $ 8.03  
        &  43.40\\
LmmdPL  
        &  53.71 $ \pm $ 1.39               &  53.25 $ \pm $ 1.27  
        &  58.71 $ \pm $ 1.42               &  \underline{48.76} $ \pm $ 0.80  
        &  61.50 $ \pm $ 3.35               &  49.72 $ \pm $ 0.81  
        &  54.28\\
DCORAL*\cite{DCORAL}    
        &  42.67 $ \pm $ 8.41               &  54.84 $ \pm $ 8.46  
        &  41.85 $ \pm $ 8.07               &  46.42 $ \pm $ 7.91  
        &  62.16 $ \pm $ 7.18               &  47.14 $ \pm $ 7.72  
        &  45.85\\
DDA*\cite{DDA}    
        &  45.38 $ \pm $ 8.51               &  45.87 $ \pm $ 8.97  
        &  54.39 $ \pm $ 4.98               &  52.53 $ \pm $ 7.82  
        &  48.46 $ \pm $ 8.62               &  44.06 $ \pm $ 8.24  
        &  46.78\\
PrPL*\cite{Li_Liang_2023}    
        &  53.40 $ \pm $ 0.86               &  54.05 $ \pm $ 2.12  
        &  \underline{59.89} $ \pm $ 0.19   &  46.97 $ \pm $ 1.60  
        &  59.39 $ \pm $ 3.08               &  \underline{50.38} $ \pm $ 1.92  
        &  54.01\\
CddPL   
        &  \underline{56.35} $ \pm $ 1.10   &  \underline{54.46} $ \pm $ 1.19           
        &  57.95 $ \pm $ 2.65               &  47.93 $ \pm $ 2.41  
        &  \underline{63.18} $ \pm $ 4.84   &  49.76 $ \pm $ 0.92  
        &  \underline{54.93}\\
BLFBA*\cite{BLFBA}    
        &  49.58 $ \pm $ 8.59               &  41.42 $ \pm $ 8.58  
        &  52.12 $ \pm $ 8.36               &  45.56 $ \pm $ 8.11  
        &  46.34 $ \pm $ 8.81               &  45.26 $ \pm $ 7.63  
        &  46.71\\
McdPL   
        &  \textbf{57.40 $ \pm $ 1.78}      &  \textbf{58.90 $ \pm $ 1.40}  
        &  \textbf{60.77 $ \pm $ 1.64}      &  \textbf{54.07 $ \pm $ 5.12}
        &  \textbf{68.83 $ \pm $ 5.74}      &  \textbf{53.42 $ \pm $ 1.50}  
        &  \textbf{58.90}\\
\midrule 
        & \textcolor{orange}{$\uparrow$ 1.05} 
        & \textcolor{orange}{$\uparrow$ 4.34}
        & \textcolor{orange}{$\uparrow$ 0.88} 
        & \textcolor{orange}{$\uparrow$ 5.31}
        & \textcolor{orange}{$\uparrow$ 5.65} 
        & \textcolor{orange}{$\uparrow$ 3.66}
        & \textcolor{orange}{$\uparrow$ 3.97}\\
%\multicolumn{2}{l}{\textbf{PADDA}} & \textbf{93.06/05.12}  \\
\midrule
\bottomrule
\end{tabular}
}
\end{center}
\end{table*}

\begin{table*}[ht]
\begin{center}
\caption{The accuracy and Standard-Deviations of the various EEG emotion recognition Methods in Cross-corpus Cross-subjects Single-session leave-one-subject-out Cross-valuation, the results are shows as (Accuracy\% $\pm$ Standard-Deviation\%). The database training combination is represented as (Source Domain $\to$ Target Domain). \textcolor{orange}{$\uparrow\%$} denotes the improvement in accuracy between the best performance and the second-best performance. Here, the model results reproduced by us are indicated by '*'.}
\label{tab:independent_validation_session_1}
\scalebox{0.8}{
\color{black}
\begin{tabular}{lccccccc}%\underline{}
\bottomrule
\midrule
Methods & SEED$\to$SEED-IV                  & SEED$\to$SEED-V 
        & SEED-IV$\to$SEED                  & SEED-IV$\to$SEED-V
        & SEED-V$\to$SEED                   & SEED-V$\to$SEED-IV
        & Ave\_Acc\\  
\midrule
DANN*\cite{DANN}    
        &  40.19 $ \pm $ 8.09               &  40.37 $ \pm $ 7.69          
        &  42.85 $ \pm $ 7.51               &  26.73 $ \pm $ 7.88  
        &  37.26 $ \pm $ 6.58               &  37.77 $ \pm $ 8.06          
        &  37.52\\
LeNet*\cite{LeNet}    
        &  42.03 $ \pm $ 8.64               &  35.22 $ \pm $ 8.56          
        &  51.91 $ \pm $ 8.20               &  32.42 $ \pm $ 7.13  
        &  42.86 $ \pm $ 7.34               &  35.24 $ \pm $ 7.83          
        &  39.94\\
LmmdPL  
        &  44.08 $ \pm $ 11.3              &  46.88 $ \pm $ 15.7   
        &  \textbf{57.09 $ \pm $ 12.9}      &  56.91 $ \pm $ 10.2  
        &  50.47 $ \pm $ 11.7               &  47.07 $ \pm $ 12.1 
        &  50.42\\
DCORAL*\cite{DCORAL}    
        &  41.67 $ \pm $ 7.94               &  42.30 $ \pm $ 8.52          
        &  46.95 $ \pm $ 7.78               &  35.59 $ \pm $ 7.76  
        &  37.73 $ \pm $ 7.71               &  36.12 $ \pm $ 6.93          
        &  40.06\\
DDA*\cite{DDA}    
        &  35.85 $ \pm $ 8.47               &  35.03 $ \pm $ 8.09          
        &  \underline{57.05} $ \pm $ 8.08   &  32.32 $ \pm $ 8.95  
        &  43.63 $ \pm $ 7.91               &  34.03 $ \pm $ 7.89          
        &  36.17\\
PrPL*\cite{Li_Liang_2023}    
        &  43.13 $ \pm $ 2.98               &  \underline{49.82} $ \pm $ 10.5          
        &  55.75 $ \pm $ 14.3              &  55.78 $ \pm $ 11.9  
        &  51.07 $ \pm $ 4.16               &  45.19 $ \pm $ 8.22          
        &  50.12\\
CddPL   
        &  \underline{44.29} $ \pm $ 3.09   &  47.80 $ \pm $ 16.8   
        &  53.35 $ \pm $ 4.45               &  \underline{60.07} $ \pm $ 13.6  
        &  \underline{51.14} $ \pm $ 5.77   &  \underline{50.34} $ \pm $ 8.84   
        &  \underline{51.17}\\
BLFBA*\cite{BLFBA}    
        &  44.23 $ \pm $ 6.89               &  49.80 $ \pm $ 8.10          
        &  51.96 $ \pm $ 7.58               &  40.17 $ \pm $ 4.49  
        &  47.36 $ \pm $ 7.29               &  47.19 $ \pm $ 5.44          
        &  50.12\\
McdPL   
        &  \textbf{56.29 $ \pm $ 6.44}      &  \textbf{57.87 $ \pm $ 8.20}  
        &  53.64 $ \pm $ 4.95               &  \textbf{60.87 $ \pm $ 4.99}
        &  \textbf{55.32 $ \pm $ 6.20}      &  \textbf{53.79 $ \pm $ 3.98}  
        &  \textbf{56.30}\\
\midrule 
        & \textcolor{orange}{$\uparrow$ 12.00} 
        & \textcolor{orange}{$\uparrow$ 8.05}
        & \textcolor{orange}{$\uparrow$ 1.34} 
        & \textcolor{orange}{$\uparrow$ 0.80}
        & \textcolor{orange}{$\uparrow$ 4.22} 
        & \textcolor{orange}{$\uparrow$ 3.45}
        & \textcolor{orange}{$\uparrow$ 5.13}\\
%\multicolumn{2}{l}{\textbf{PADDA}} & \textbf{93.06/05.12}  \\
\midrule
\bottomrule
\end{tabular}
}
\end{center}
\end{table*}

\begin{table*}[ht]
\begin{center}
\caption{The accuracy and Standard-Deviations of the various EEG emotion recognition Methods in Cross-corpus Cross-subjects Cross-session leave-one-subject-out Cross-valuation, the results are shows as (Accuracy\% $\pm$ Standard-Deviation\%). The database training combination is represented as (Source Domain $\to$ Target Domain). \textcolor{orange}{$\uparrow\%$} denotes the improvement in accuracy between the best performance and the second-best performance. Here, the model results reproduced by us are indicated by '*'.}
\label{tab:inpandent valation allsession}
\scalebox{0.8}{
\color{black}
\begin{tabular}{lccccccc}%\underline{}
\bottomrule
\midrule
Methods & SEED$\to$SEED-IV                  & SEED$\to$SEED-V 
        & SEED-IV$\to$SEED                  & SEED-IV$\to$SEED-V
        & SEED-V$\to$SEED                   & SEED-V$\to$SEED-IV
        &Ave\_Acc\\  
\midrule
DANN*\cite{DANN}    
        &  44.10 $ \pm $ 8.69               &  40.10 $ \pm $ 8.49          
        &  47.32 $ \pm $ 8.22               &  41.25 $ \pm $ 7.75  
        &  48.72 $ \pm $ 8.52               &  34.35 $ \pm $ 8.57          
        &  42.64\\
LeNet*\cite{LeNet}    
        &  46.01 $ \pm $ 8.56               &  41.31 $ \pm $ 6.66          
        &  47.01 $ \pm $ 8.65               &  39.10 $ \pm $ 7.68  
        &  45.03 $ \pm $ 8.19               &  40.20 $ \pm $ 7.79          
        &  43.11\\
LmmdPL  
        &  49.17 $ \pm $ 2.06               &  47.65 $ \pm $ 11.5  
        &  \underline{56.52} $ \pm $ 3.97   &  \underline{53.62} $ \pm $ 7.03
        &  \underline{49.60} $ \pm $ 5.53   &  \textbf{53.05 $ \pm $ 5.99}  
        &  \underline{51.60}\\
DCORAL*\cite{DCORAL}    
        &  43.76 $ \pm $ 8.36               &  44.20 $ \pm $ 7.62          
        &  43.24 $ \pm $ 7.89               &  44.63 $ \pm $ 7.83  
        &  44.82 $ \pm $ 7.40               &  39.95 $ \pm $ 7.15          
        &  43.43\\
DDA*\cite{DDA}    
        &  44.23 $ \pm $ 8.48               &  41.11 $ \pm $ 8.27          
        &  54.50 $ \pm $ 8.31               &  42.09 $ \pm $ 7.46  
        &  49.45 $ \pm $ 9.08               &  43.76 $ \pm $ 8.14          
        &  50.12\\
PrPL*\cite{Li_Liang_2023}    
        &  51.27 $ \pm $ 1.29               &  \underline{52.16} $ \pm $ 10.2 
        &  56.39 $ \pm $ 5.27               &  45.26 $ \pm $ 7.90
        &  48.95 $ \pm $ 2.08               &  \underline{48.13} $ \pm $ 6.73   
        &  50.36\\
CddPL   
        &  \textbf{52.89 $ \pm $ 5.78}      &  51.41 $ \pm $ 8.40  
        &  \textbf{56.55 $ \pm $ 5.53}      &  51.83 $ \pm $ 3.75
        &  46.33 $ \pm $ 5.40               &  48.05 $ \pm $ 7.36           
        &  51.18\\
BLFBA*\cite{BLFBA}    
        &  50.75 $ \pm $ 6.00               &  51.29 $ \pm $ 6.11          
        &  55.81 $ \pm $ 7.56               &  52.13 $ \pm $ 8.75  
        &  49.18 $ \pm $ 7.30               &  50.28 $ \pm $ 6.57          
        &  50.12\\
McdPL   
        &  \underline{51.35} $ \pm $ 3.08   &  \textbf{54.50 $ \pm $ 2.94}  
        &  51.37 $ \pm $ 4.67               &  \textbf{55.72 $ \pm $ 3.30}
        &  \textbf{49.80 $ \pm $ 4.21}      &  47.66 $ \pm $ 2.61 
        &  \textbf{51.73}\\
\midrule 
        & \textcolor{orange}{$\uparrow$ 1.62} 
        & \textcolor{orange}{$\uparrow$ 2.34}
        & \textcolor{orange}{$\uparrow$ 0.03} 
        & \textcolor{orange}{$\uparrow$ 2.10}
        & \textcolor{orange}{$\uparrow$ 0.20} 
        & \textcolor{orange}{$\uparrow$ 4.92}
        & \textcolor{orange}{$\uparrow$ 0.13}\\
%\multicolumn{2}{l}{\textbf{PADDA}} & \textbf{93.06/05.12}  \\
\midrule
\bottomrule
\end{tabular}
}
\end{center}
\end{table*}
%----------------------------------------------------------------------------------

\section{Experimental Results}
\label{sec:experiment resules}
\begin{CJK*}{UTF8}{gbsn}

\subsection{Emotional Dataset and Data Preprocessing}
    \indent We validated our proposed three models on three well-known public databases: SEED \cite{Wei_2015_eeg}, SEED-IV \cite{pi_2018_uns} and SEED-V \cite{Li_2021_eeg}. These databases have been widely utilized in research \cite{Cui_2020_eeg,Liu_2024_eeg,Zhang_2019_eeg,Feng_2022_eeg}. In the SEED and SEED-IV databases, a total of 15 subjects were participated in the experiments, with each subject completing three sessions on different dates. Each session in SEED consisted of 15 trials and included three emotions: negative, neutral and positive. Each session in SEED-IV consisted of 24 trials and included four emotions: happiness, sadness, fear, and neutral. In the SEED-V database, a total of 16 subjects were participated, with each completing three sessions on different dates. Each session comprised 15 trials and include five emotions: happiness, neutral, sadness, disgust and fear.
    \\ \indent To consistency and rigor in our experiments, we defined happiness as a positive emotion and sadness as a negative emotion. we retained only positive, neutral, and negative emotional states in all databases. In the SEED-IV database, samples associated with the emotion of fear were excluded, while in the SEED-V database, samples associated with he emotion of fear and disgust were excluded. During the EEG signal collection process, all databases both utilized a 62-channel ESI neuroscan system, and all data underwent uniform preprocessing procedures.We downsampled the EEG data to 200 Hz and manually removed contaminated signals, such as EMG and EOG. Subsequently, the EEG data were filtered using a band-pass filter with a range of 0.3 Hz to 50 Hz. We set a window length of 1 second, segmenting the data for each trial into non-overlapping segments of 1 second. Based on five predefined frequency bands, Delta (1-3 Hz), Theta (4-7 Hz), Alpha (8-13 Hz), Beta (14-30 Hz) and Gamma (31-50 Hz), the corresponding signals is extracted according to these five frequency bands. Then, the Differential Entropy (DE \cite{Duan_Zhu_2013}) was computed to represent the logarithm energy spectrum of each specific frequency band. For a sequence $X$ that follows a Gaussian distribution $ N(\mu,\delta^2)$, the differential entropy for the $i_{th}$ frequency band can be defined as :
    \begin{equation}
    \label{Eq:21}
    \begin{split}
        h_i(X) =& \int\limits_{-\infty}^{\infty}  {\frac{1}{\sqrt{2\pi\delta^2_i}} e^{\frac{(x-\mu)^2}{2\delta^2_i}}}\log_{}{\left (   \frac{1}{\sqrt{2\pi\delta^2_i }}e^{\frac{(x-\mu)^2}{2\delta^2_i}}\right )}dx \\=&\frac{1}{2}\log_{}{\left ( 2\pi e\delta ^2_i\right )}
    \end{split}
    \end{equation}
    \indent Thus, a complete emotional data segment contains 310 features (62 channels × 5 frequency bands). We applied a Linear Dynamic System (LDS) approach to smooth all features, which effectively leverages the temporal dependencies of emotional variations and filters out EEG components of unrelated and noise \cite{Li_2010_eeg}. Significantly, during the preprocessing of cross-validation data, there is no issue of any information leakage, as the source and target domains samples from the independent databases.

\subsection{Implementation Results}
    \indent The model's feature extractor $f(\cdot)$ and discriminator $D(\cdot)$ are constructed using Multilayer Perceptron (MLP) with ReLU activation functions. The parameters of the bilinear transformation matrix $\theta$ are initialized randomly by uniform distribution. In the model architecture, the feature extractor is designed as: input layer (310) – hidden layer 1 (128) – ReLU activation – hidden layer 2 (64) – ReLU activation – feature output layer (64). The domain discriminator is designed as: feature input layer (64) – hidden layer 1 (128) – ReLU activation – dropout layer – hidden layer 2 (64) – Sigmoid activation (1) – output layer (1). In the LmmdPL and CddPL model, Since there is only one classifier, We randomly initialize the parameters and utilize the Rmsprop optimizer to optimize the model. In the McdPL model, the two classifiers utilize the Adam optimizer and RMSprop optimizer, respectively, which demonstrating better performance compared to other classical optimizers. The learning rate is set to 1e-3, with a training epoch size of 300 and a batch size of 256. To avoid overfitting, L2 regularization with a weight of \(1 \times 10^{-5}\) is employed during model training. All models are executed under the following configuration: NVIDIA GeForce RTX 3090, CUDA=11.6, PyTorch =1.12.1. Significantly, the label information of the target domain is completely missing in the training process, consistent with previously employed EEG emotion recognition models utilizing transfer learning \cite{Lis_2022_eeg,Zhong_2022_eeg,Wu_2022_eeg}.

%----------------------------------------------------------------------------
\begin{figure*}[ht]
\centering
\subfloat{\includegraphics[width=1\textwidth]{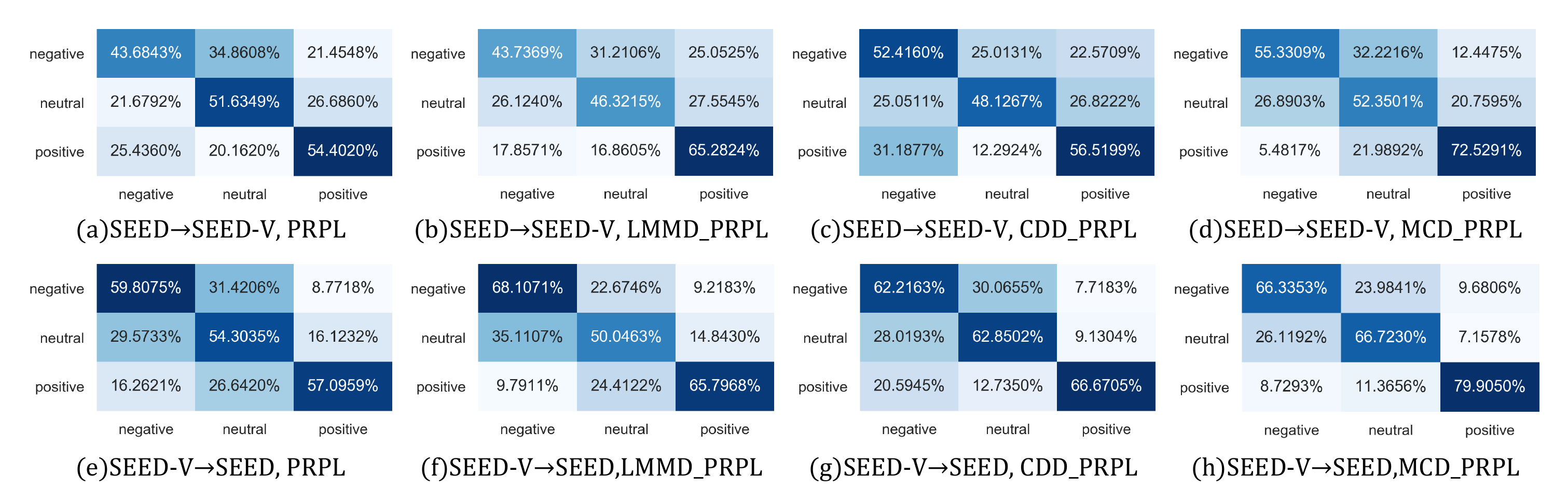}}
\caption{
    The confusion matrices for PrPL, LmmdPL, CddPL and McdPL during cross-corpus training. In matrices (a)$\sim$(d), SEED is utilized as the source domain and SEED-V as the target domain, denoted as SEED$\to$SEED-V. Conversely, matrices (e)$\sim$(h) use SEED-V as the source domain and SEED as the target domain, represented as SEED-V$\to$SEED. In each confusion matrix view, the horizontal axis represents the predicted labels, while the vertical axis represents the true labels.
    }
\label{fig:confusion_matrix}
\end{figure*}
%---------------------------------------------------------------------------

\subsection{Experiment Protocols}
\label{experiment protocols}
    \indent To comprehensively evaluate the robustness and stability of the models, we randomly selected source domain data and target domain data from the three databases, denoted as (Source Domain$\to$Target Domain). It produce six different cross-corpus training combinations: SEED$\to$SEED-IV, SEED$\to$SEED-V, SEED-IV$\to$SEED, SEED-IV$\to$SEED-V, SEED-V$\to$SEED and SEED-V$\to$SEED-IV, respectively. Additionally, we employed four different validation protocols.
    \textbf{(1) Cross-corpus Cross-subjects Single-session hold-out-valuation}. In single-database EEG emotion recognition tasks, utilizing the first session for model evaluation is the most widely adopted validation protocol \cite{WGAN_2018,Li_Qiu_2020,Zheng_Zhang_2015_eeg}. The first session of all subjects from one database was designated as the source domain data, while the first session of all subjects from another database was used as the target domain data.
    \textbf{(2) Cross-corpus Cross-subjects Cross-session hold-out-valuation}. all sessions from all subjects within one database as source domain data and the another database as target domain data. since the cross-corpus cross-subject cross-session, this evaluation protocol presents significant challenges in EEG emotion recognition tasks.
    \textbf{(3) Cross-corpus Cross-subjects Single-session leave-one-subject-out Cross-valuation}. To more rigorously validate the robustness and generalization performance of our model across databases, we will use an independent validation set for evaluation. Specifically, in one database, the first session of all subjects will be used as source domain sample data. In another database, the first session of one subject will be treated as the independent validation set, which will not participate in training or testing, and will be used to validate the overall performance of the model after training, while the first session of the remaining subjects will be used as target domain sample data. This process will continue until the first session of each subject in the target domain database has been used as the independent validation set. We will perform 15-fold (SEED,SEED-IV) or 16-fold (SEED-V) cross-validation, and the results will be averaged.
    \textbf{(4) Cross-corpus Cross-subjects Cross-session leave-one-subject-out Cross-valuation}. Specifically, all sessions of all subjects from one database are treated as source domain sample data, while all sessions of one subject from another database serve as the independent validation set, and the sessions of the remaining subjects are used as target domain sample data. This process will continue until each subject in the target domain database has been used as the independent validation set. We will perform 15-fold (SEED, SEED-IV) or 16-fold (SEED-V) cross-validation, and the results will be averaged. This evaluation protocol imposes more stringent requirements on the model.

\subsection{Cross-corpus Cross-subjects Single-session hold-out-valuation}
    \indent The training results are shown in Tab.\ref{tab:test_session_1}, all results are represented as (Accuracy\% $\pm$ Standard-Deviation\%), State-of-The-Art (SOTA) performances are shown in black bold, and the second-best performance are underlined. The results show that the LmmdPL, CddPL and McdPL Methods are superior to the baseline models. Notably, the McdPL achieves SOTA performance across all six combinations of training cross-corpus. CddPL achieves the second-best accuracy in three out of six training cross-corpus combinations (SEED$\to$SEED-IV, SEED-IV$\to$SEED-V, SEED-V$\to$SEED-IV), while LmmdPL alse gets three (SEED$\to$SEED-V, SEED-IV$\to$SEED, SEED-V$\to$SEED). It is worth noting that the average accuracy of LmmdPL, CddPL and McdPL models is 53.18\%, 54.17\% and 58.93\%, respectively. Compared with the second-best model (CddPL), the average accuracy of McdPL is improved by 4.76\%. These results demonstrate that McdPL achieve superior fine-grained feature alignment and EEG emotion recognition performance.

\subsection{Cross-corpus Cross-subjects Cross-session hold-out-valuation}
    \indent This validation protocol presents significant challenges in EEG emotion recognition tasks. The experimental results presented in Tab.\ref{tab:test_all_session} demonstrate that the McdPL model consistently achieves SOTA performance across all six training cross-corpus combinations. Notably, the cross-corpus evaluation from SEED-V $\to$ SEED yields the highest accuracy of 68.53\%. Among the comparative methods, CddPL achieves second-best performance in three cross-corpus training combinations, while LmmdPL attains this position in one configuration. The average accuracy analysis reveals comparable performance levels between PrPL (54.01\%), LmmdPL (54.28\%), and CddPL (54.93\%) methods. It is worth noting that the McdPL framework achieves a mean accuracy of 58.90\%, representing a 3.97\% improvement over the second-best performing method (CddPL). These results demonstrate the superior generalization capabilities and potential of the proposed McdPL architecture.

\subsection{Cross-corpus Cross-subjects Single-session leave-one-subject-out Cross-valuation}
    \indent Since the independent validation set excluded from model training, this experimental protocols eliminates potential bias contamination while presenting enhanced challenges for model generalization. As shown in Tab.\ref{tab:independent_validation_session_1} (with SOTA performance highlighted in black bold), the LmmdPL model achieves optimal performance on the SEED-IV$\to$SEED cross-corpus evaluation combination, attaining an accuracy of 57.09\%. Notably, the McdPL framework achieve SOTA performances across all remaining cross-corpus validation combinations, achieving a mean accuracy of 56.3\% that demonstrates a 5.13\% improvement over secondary methods (CddPL). Collectively, these results demonstrate the superior performance and cross-corpus generalization capabilities of the McdPL architecture compared to alternative approaches. 
    
\subsection{Cross-corpus Cross-subjects Cross-session leave-one-subject-out Cross-valuation}
    \indent The performance results of the model are shown in Tab.\ref{tab:inpandent valation allsession}. The McdPL framework demonstrates superior accuracy in the SEED$\to$SEED-V, SEED-IV$\to$SEED-V and SEED-V$\to$SEED cross-corpus evaluations combinations, and achieve the SOTA performance in three of the six cross-corpus training combinations, accuracy are 54.50\%, 55.72\% and 49.80\%, respectively. The CddPL model achieved the superior accuracy in the SEED$\to$SEED-IV and SEED-IV$\to$SEED cross-corpus combinations, with an accuracy of 52.89\% and 56.55\%, respectively. Meanwhile, the LmmdPL model achieved the superior accuracy in the SEED-V$\to$SEED-IV combination, with an accuracy of 53.05\%. Overall, the McdPL model continues to show the most potential, with an average accuracy of 51.73\% percent across all cross-corpus combinations, slightly outperforming the other models.
    
\end{CJK*}

%----------------------------------------------------------------------------
\begin{figure*}[ht]
\centering
\subfloat{\includegraphics[width=1\textwidth]{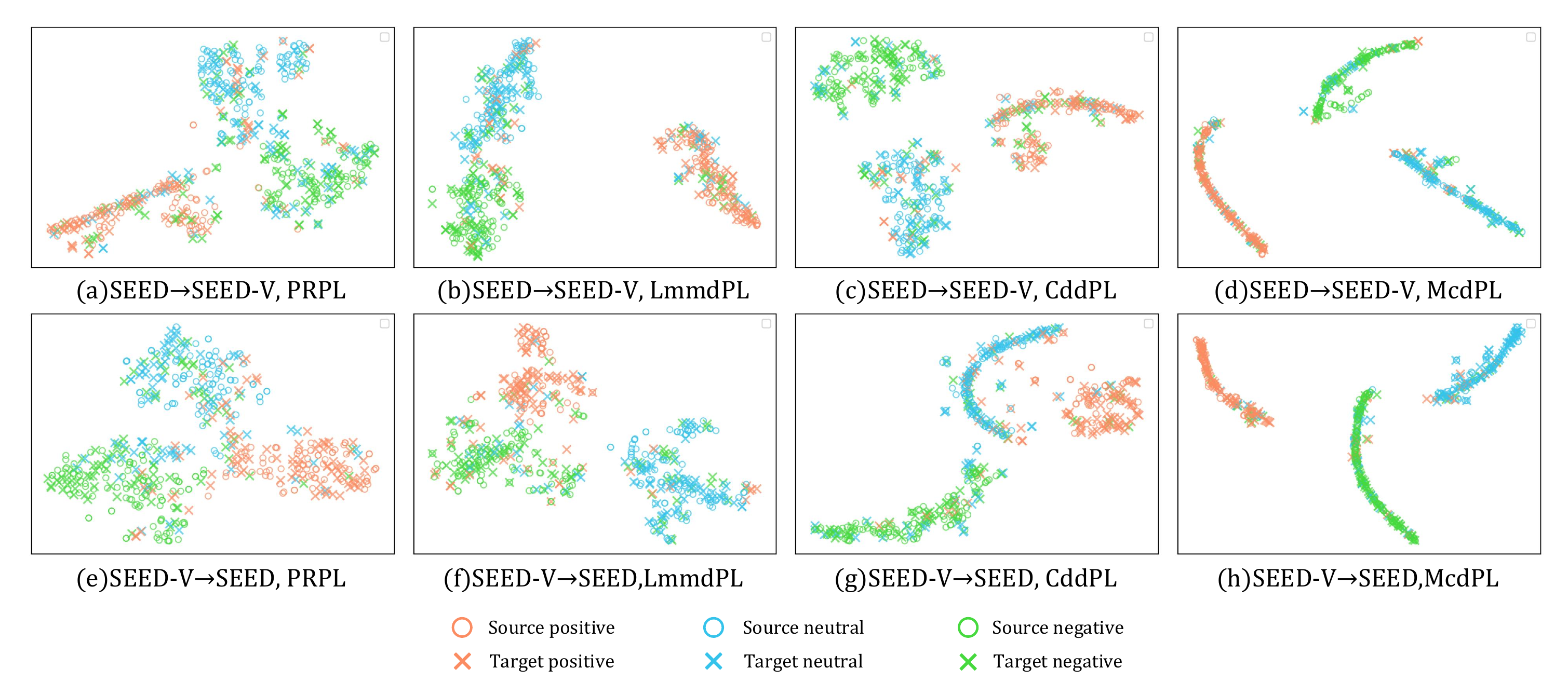}}
\caption{
    Displays the T-SNE visualization of the sample features learned by the model (PrPL,LmmdPL,CddPL,McdPL) in both the source and target domains.(a)$\sim$(d) utilize SEED as the source domain and SEED-V as the target domain, represented as SEED$\to$SEED-V. Conversely, (e)$\sim$(h) utilize SEED-V as the source domain and SEED as the target domain, denoted as SEED-V$\to$SEED. Different colors in the visualization correspond to different categories.
    }
\label{fig:Vision}
\end{figure*}
%---------------------------------------------------------------------------

%---------------------------------------------------------------------------
\begin{table*}[t]
\begin{center}
\caption{The results of the McdPL model adding different proportions of noisy labels to the source domain in the pairwise learning and pointwise learning strategies, expressed as (Accuracy$\% \pm$Standard-Deviation\%). \textcolor{orange}{$\downarrow\%$} denotes the difference between the performances.
}
\label{tab:noisy labels}
\scalebox{0.97}{
\color{black}
\begin{tabular}{llcccccc}
\toprule
\toprule
  & Strategy  &0\%  &10\%  &20\%  &30\%  &40\%  &40\%-10\%\\
\midrule
SEED$\to$SEED-IV
&Pairwise   & 57.40 $ \pm $ 1.78             & 55.51 $ \pm $ 2.07 
            & 52.16 $ \pm $ 6.18             & 50.03 $ \pm $ 4.93 
            & 51.95 $ \pm $ 0.93
            & \textcolor{orange}{$\downarrow$ 3.56} \\
&Pointwise  & 51.45 $ \pm $ 1.37             & 48.82 $ \pm $ 1.77 
            & 45.13 $ \pm $ 5.68             & 43.09 $ \pm $ 5.95 
            & 41.67 $ \pm $ 4.76
            & \textcolor{pink}{$\downarrow$ 7.15}\\
\midrule 
SEED$\to$SEED-V 
&Pairwise   & 58.90 $ \pm $ 1.40             & 53.06 $ \pm $ 0.52 
            & 52.02 $ \pm $ 6.15             & 52.23 $ \pm $ 0.33 
            & 51.51 $ \pm $ 7.72
            & \textcolor{orange}{$\downarrow$ 1.52}\\
&Pointwise  & 52.64 $ \pm $ 6.68             & 48.90 $ \pm $ 4.45  
            & 47.84 $ \pm $ 5.85             & 46.03 $ \pm $ 7.52 
            & 44.85 $ \pm $ 4.16 
            & \textcolor{pink}{$\downarrow$ 4.32}\\
\midrule 
SEED-IV$\to$SEED 
&Pairwise   & 60.77 $ \pm $ 1.64             & 60.48 $ \pm $ 9.30 
            & 60.40 $ \pm $ 7.44             & 60.20 $ \pm $ 8.89 
            & 59.96 $ \pm $ 9.61
            & \textcolor{orange}{$\downarrow$ 0.52}\\
&Pointwise  & 56.83 $ \pm $ 5.77             & 54.25 $ \pm $ 6.57 
            & 53.01 $ \pm $ 6.51             & 49.96 $ \pm $ 7.36 
            & 48.13 $ \pm $ 7.03
            & \textcolor{pink}{$\downarrow$ 6.12}\\
\midrule 
SEED-IV$\to$SEED-V 
&Pairwise   & 54.07 $ \pm $ 5.12             & 52.01 $ \pm $ 7.03 
            & 50.56 $ \pm $ 6.09             & 49.34 $ \pm $ 6.09 
            & 48.62 $ \pm $ 5.37 
            & \textcolor{orange}{$\downarrow$ 3.39}\\
&Pointwise  & 51.16 $ \pm $ 5.72             & 48.39 $ \pm $ 4.16 
            & 45.76 $ \pm $ 1.95             & 44.34 $ \pm $ 3.07 
            & 43.81 $ \pm $ 3.01 
            & \textcolor{pink}{$\downarrow$ 4.58}\\
\midrule 
SEED-V$\to$SEED 
&Pairwise   & 68.83 $ \pm $ 5.74             & 67.97 $ \pm $ 1.60 
            & 66.16 $ \pm $ 0.70             & 65.22 $ \pm $ 3.03 
            & 64.57 $ \pm $ 3.11 
            & \textcolor{orange}{$\downarrow$ 3.40}\\
&Pointwise  & 65.11 $ \pm $ 7.51             & 55.92 $ \pm $ 6.28
            & 52.99 $ \pm $ 5.90             & 51.09 $ \pm $ 6.57 
            & 49.51 $ \pm $ 4.68 
            & \textcolor{pink}{$\downarrow$ 6.41}\\
\midrule 
SEED-V$\to$SEED-IV 
&Pairwise   & 53.42 $ \pm $ 1.50             & 53.30 $ \pm $ 0.54 
            & 51.79 $ \pm $ 1.04             & 50.10 $ \pm $ 5.38 
            & 48.76 $ \pm $ 5.04 
            & \textcolor{orange}{$\downarrow$ 4.45}\\
&Pointwise  & 45.25 $ \pm $ 2.45             & 45.08 $ \pm $ 4.13 
            & 43.95 $ \pm $ 4.30             & 41.57 $ \pm $ 2.76 
            & 39.72 $ \pm $ 3.46 
            & \textcolor{pink}{$\downarrow$ 5.36}\\
\midrule 
Mean\_Acc 
&Pairwise   & 58.90         & 57.04 
            & 55.52         & 54.52 
            & 54.22 
            & \textcolor{orange}{$\downarrow$ 2.82}\\
&Pointwise  & 53.74         & 50.22 
            & 48.11         & 46.01
            & 44.57 
            & \textcolor{pink}{$\downarrow$ 5.65}\\
Difference in Mean\_Acc
&           & \textcolor{orange}{$\downarrow$ 5.16} 
            & \textcolor{orange}{$\downarrow$ 6.82}
            & \textcolor{orange}{$\downarrow$ 7.41}
            & \textcolor{orange}{$\downarrow$ 8.51}
            & \textcolor{orange}{$\downarrow$ 9.65}
            & \\
\toprule
\bottomrule
\end{tabular}
}
\end{center}
\end{table*}
%---------------------------------------------------------------------------

\section{Discussion and Conclusion} 
\label{sec:Discussion_and_Conclusion}
\begin{CJK*}{UTF8}{gbsn}

\subsection{Confusion Matrix}
    \indent As shown in Fig.\ref{fig:confusion_matrix}, we present the confusion matrices for the models (PrPL, LmmdPL, CddPL and McdPL) in SEED$\to$SEED-V and SEED-V$\to$SEED. In Fig.\ref{fig:confusion_matrix}.(a)$\sim$(d), It is shown that the McdPL model achieved the best performance in all models, with accuracy of 55.33\%, 52.25\% and 72.53\% for correctly identifying negative, neutral and positive emotion samples, respectively. Additionally, in the negative and positive emotion categories, the CddPL model achieved the second-best recognition performance, with accuracies of 52.42\% and 56.52\%. The PrPL model demonstrated the second-SOTA performance in the neutral emotion category, with an accuracy of 51.63\%. In Fig.\ref{fig:confusion_matrix}.(e)$\sim$(h), It is shown that the McdPL model achieved accuracy of 66.72\% and 79.91\% for correctly identifying neutral and positive emotion samples, respectively. The LmmdPL model recorded an accuracy of 68.11\% for negative emotion samples, while the McdPL model achieved 66.34\%, indicating the latter's second-best performance. Overall, The McdPL model still shows the SOTA performance of emotion recognition, while the LmmdPL and CddPL models slightly outperform the PrPL model.

\subsection{Visualization of Learned Representation}
    \indent To provide a more intuitive comparison of the feature representation effects of different models, we used T-distributed Stochastic Neighbor Embedding (T-SNE \cite{Maaten_Hinton_2008}) algorithm to visualize the feature samples and interactive features in the source and target domains of the PrPL, LmmdPL, CddPL and McdPL models. As shown in Fig.\ref{fig:Vision}. This visualization demonstrate each model’s ability to learn discriminative features and achieve class differentiation. Specifically, we randomly selected 256 samples each from the source and target domains to visualize the learned feature representations, respectively. In cross-corpus training combinations SEED→SEED-V and SEED-V→SEED, which are significant performance differences across models. In the feature distribution of the PrPL model (Fig.\ref{fig:Vision}.(a)(e)), the distribution of each category is the most dispersed, with blurred boundaries between classes, indicating the feature differentiation of samples is weak. In contrast, the feature distributions in the LmmdPL (Fig.\ref{fig:Vision}.(b)(f)) and CddPL (Fig.\ref{fig:Vision}.(c)(g)) models show increasing Clustering ability, with relatively clear boundaries forming between categories, reflecting stronger feature alignment potential. Significantly, the feature distribution obtained with the McdPL model (Fig.\ref{fig:Vision}.(d)(h)) is the most compact, with the clearest boundaries between categories. This dense and distinct distribution demonstrate the McdPL model's strong cross-domain feature alignment capability and excellent classification performance, giving it a significant advantage in emotion recognition tasks.
    
\subsection{Effect of Noisy Labels}
    \indent We adopt a pairwise learning strategy to effectively overcome the problem of label noise. In order to verify the robustness of the model under the pairwise learning and pointwise learning strategies and its dependence on the sample labels, we train the model using the source domain samples with added noise. Specifically, we introduce noise into the real label of the source domain in a controlled manner and evaluate the model using a sample of the target domain. We generate a large number of random labels and replace 10\%, 20\%, 30\% and 40\% of the source domain sample labels with these random labels. It is worth noting that during the training process, the label information of the target domain is completely lost. 
    \\ \indent We adopt the McdPL model as the baseline model and the results are shown in Tab.\ref{tab:noisy labels}. The introduction of noise in the source domain labels in the pairwise learning strategy results in a slight degradation of the model performance. When the noise level is 10\%, 20\%, 30\% and 40\%, the average accuracy is 57.04\%, 55.52\%, 54.52\% and 54.22\%, respectively, which is only reduced by 2.82\%.
    However, the introduction of noise in the source domain labels in the pointwise learning strategy leads to a significant degradation in the model performance. When the noise level is 10\%, 20\%, 30\% and 40\%, the average accuracy is 50.22\%, 48.11\%, 46.01\% and 44.57\%, respectively, which is reduced by 5.65\%.
    It is worth noting that compared with the pairwise learning strategy, the performance of the model is affected more by the increase of the proportion of label noise in the pointwise learning strategy. Without adding label noise (0\%), the average accuracy is 58.90\% and 53.74\%, with a difference of 5.16\%. However, when the proportion of label noise reaches 40\%, the average accuracy is 54.22\% and 44.57\%, with a difference of 9.65\%.
    \\ \indent These results show that increasing the proportion of noise in the source domain sample has a limited effect on model performance. Overall, our proposed McdPL model based on pairwise learning has excellent robustness and reliability, and has a high tolerance for noise labels.

%---------------------------------------------------------------------------
\begin{table*}[t]
\begin{center}
\caption{The ablation experiment of our proposed model, expressed as (Accuracy$\% \pm$Standard-Deviation\%). \textcolor{orange}{$\downarrow\%$} is denoted as the gap from the best performance.
}
\label{tab:Ablation Experiment}
\scalebox{0.91}{
\color{black}
\begin{tabular}{llcc|cc|cc}
\bottomrule
\midrule
Strategy  &  &LmmdPL &  &CddPL & &McdPL &  \\
\midrule
w/o pairwise learning on the target 
    & 
    & 54.18 $ \pm $ 4.22     & \textcolor{orange}{$\downarrow$ 4.09}
    & 54.19 $ \pm $ 2.30     & \textcolor{orange}{$\downarrow$ 2.09}
    & 65.11 $ \pm $ 7.51     & \textcolor{orange}{$\downarrow$ 3.64}\\
w/o pairwise learning on the source and target 
    &
    & 44.55 $ \pm $ 6.87     & \textcolor{orange}{$\downarrow$ 13.7}
    & 46.36 $ \pm $ 1.18     & \textcolor{orange}{$\downarrow$ 9.92}
    & 63.66 $ \pm $ 5.09     & \textcolor{orange}{$\downarrow$ 10.3}\\
w/o prototypical representation 
    &
    & 54.46 $ \pm $ 3.66     & \textcolor{orange}{$\downarrow$ 3.81}
    & 49.95 $ \pm $ 3.51     & \textcolor{orange}{$\downarrow$ 6.33}
    & 58.62 $ \pm $ 5.39     & \textcolor{orange}{$\downarrow$ 10.1}\\
w/o the bilinear transformation matrix $\theta$ in Sec.\ref{sec:Pairwise Learning}
    &
    & 51.82 $ \pm $ 4.20     & \textcolor{orange}{$\downarrow$ 6.45}
    & 47.01 $ \pm $ 1.81     & \textcolor{orange}{$\downarrow$ 9.27}
    & 66.72 $ \pm $ 9.98     & \textcolor{orange}{$\downarrow$ 2.03}\\
w/o feature discriminator 
    &
    & 53.21 $ \pm $ 3.33     & \textcolor{orange}{$\downarrow$ 5.06}
    & 51.22 $ \pm $ 2.90     & \textcolor{orange}{$\downarrow$ 5.06}
    & 59.74 $ \pm $ 10.6     & \textcolor{orange}{$\downarrow$ 9.01}\\
\midrule
A single pairwise learning classifier 
    &
    & 58.27 $ \pm $ 4.57     & \textcolor{orange}{---}
    & 56.28 $ \pm $ 2.92     & \textcolor{orange}{---}
    & 55.54 $ \pm $ 4.20     & \textcolor{orange}{$\downarrow$ 13.2}\\
w/o Maximize Classifiers Discrepancy(step.2 in McdPL) 
    &
    & ---                    &
    & ---                    &
    & 59.81 $ \pm $ 1.79     & \textcolor{orange}{$\downarrow$ 8.94}\\
w/o Minimize Features Distribution(step.3 in McdPL) 
    &
    & ---                    &
    & ---                    &
    & 54.02 $ \pm $ 7.25     & \textcolor{orange}{$\downarrow$ 14.7}\\
w/o step.2 and step.3 in McdPL 
    &
    & ---                    &
    & ---                    &
    & 58.38 $ \pm $ 7.11     & \textcolor{orange}{$\downarrow$ 10.4}\\
\midrule
$\gamma$ value in LmmdPL (Eq.\ref{Eq:13}) and CddPL (Eq.\ref{Eq:15}) 
    & 0.2
    & 58.18 $ \pm $ 4.15     & \textcolor{orange}{$\downarrow$ 0.09}
    & 54.84 $ \pm $ 4.35     & \textcolor{orange}{$\downarrow$ 1.44}
    & ---                    &\\
    &0.5
    & 58.27 $ \pm $ 4.57     & \textcolor{orange}{---}
    & 51.51 $ \pm $ 2.91     & \textcolor{orange}{$\downarrow$ 4.77}
    & ---                    &\\
    & 1.0
    & 54.78 $ \pm $ 2.65     & \textcolor{orange}{$\downarrow$ 3.49}
    & 56.28 $ \pm $ 2.92     & \textcolor{orange}{---}
    & ---                    &\\
    & 1.5
    & 56.45 $ \pm $ 3.20     & \textcolor{orange}{$\downarrow$ 1.82}
    & 54.39 $ \pm $ 2.89     & \textcolor{orange}{$\downarrow$ 1.89}
    & ---                    &\\
    & 2.0
    & 54.70 $ \pm $ 4.13     & \textcolor{orange}{$\downarrow$ 3.57}
    & 50.52 $ \pm $ 2.50     & \textcolor{orange}{$\downarrow$ 5.76}
    & ---                    &\\
\midrule
Ours 
    &
    & \textbf{58.27 $ \pm $ 4.57}     &
    & \textbf{56.28 $ \pm $ 2.92}     &
    & \textbf{68.75 $ \pm $ 9.35}      &\\
\midrule
\bottomrule
\end{tabular}
}
\end{center}
\end{table*}
%---------------------------------------------------------------------------

\subsection{Ablation Experiment}
    \indent We systematically explored the effectiveness of different components in the proposed model through ablation experiment and examined the corresponding contributions to the overall performance, and the results  are shown in Tab.\ref{tab:Ablation Experiment}. Pairwise learning strategies have a better tolerance for the inevitable label noise in emotional EEG data. Therefore, when we removed \textit{the pairwise learning strategy from the target domain}, the accuracy of the proposed LmmdPL, CddPL and McdPL were 54.18\%, 54.19\% and 65.11\%, respectively, and the performances decreased by 4.09\%, 2.09\% and 3.64\%, respectively. Additionally, when we remove \textit{the pairwise learning strategy from both the source and target domain}, the performance of the three proposed models is significantly decreased by 13.7\%, 9.92\%, and 10.3\%, respectively. These results show that the pairwise learning strategy effectively improves the performance and effectiveness of the model.
    When \textit{the prototype representation} module is removed, the performance of CddPL and McdPL is significantly decreased by 6.33\% and 10.1\%, respectively. This indicates that the extraction of prototype representation makes a significant contribution to the model. 
    In addition, the introduction of \textit{the bilinear transformation matrix $\theta$} proposed in Sec.\ref{sec:Pairwise Learning} can improve the model performance, and the recognition accuracy in CddPL is improved by 9.27\% (from 47.01\% to 56.28\%).  
    Experiments show that the introduction of domain adversarial training can greatly enhance the emotion recognition performance on the target domain. When the model removes \textit{the discriminator loss function}, the performance of all three models decreases. The recognition accuracy of LmmdPL, CddPL and McdPL is 53.21\%, 51.22\% and 59.74\%, and the performance is decreased by 5.06\%, 5.06\% and 9.01\%, respectively. This significant decrease indicates the significant impact of individual differences issues on model performance and highlights the great potential of transfer learning in affective brain-computer interface (aBCI) applications.
    \\ \indent In the MCdPL model, we adopted two classifiers based on pairwise learning and used three training steps to train the model. When we removed \textit{training step.2} (Maximize classification Discrepancy), the model performance decreased by 8.94\% (from 68.75\% to 59.81\%). When we removed \textit{training step.3} (Minimize distribution difference), the model performance decreased by 14.7\% (from 68.75\% to 54.02\%). This further demonstrates the contribution of the three training steps to the processing ability of samples located near the decision boundary. In addition, we remove \textit{step.2 and step.3 in McdPL}, and use the traditional method to train the model. The results shown that the accuracy is 58.38\%, and the model performance is significantly decreased by 10.4\%.
    \\ \indent Furthermore, we tested the value of the hyperparameter $\gamma$ in the loss function (Eq.\ref{Eq:13}) of the LmmdPL model. The results show that when the value of $\gamma$ is 0.5, the model demonstrated the optimal performance, with the accuracy reaching 58.27\%. Similarly, we tested the value of $\gamma$ in the loss function (Eq.\ref{Eq:15}) of the CddPL model. The results show that when the value of $\gamma$ is 1, the model exhibits the optimal performance.

\subsection{Conclusion}
    \indent This work was proposed three cross-corpus transfer learning Methods based on pairwise learning to enhance feature alignment between source and target domain samples, thereby improving cross-corpus emotion recognition performance. In the domain adaptation approach, we proposed the LmmdPL and CddPL Methods to achieve finer-grained alignment of samples in the feature space. For rule-domain adaptation, we proposed the McdPL framework, which designs two distinct classifiers and incorporates three specialized training steps to explore a more suitable feature space for aligning sample features. The models were comprehensively evaluated on the SEED, SEED-IV and SEED-V databases. In addition, the stability and generalization of the model under various experimental settings were thoroughly validated. The experimental results show that Lmmd, CddPL and McdPL models have achieved excellent performance. Among them, the McdPL model achieves the optimal performance and is superior in dealing with the individual differences and noisy labeling problems in aBCI systems, which provides a promising solution for cross-corpus emotion recognition. In future studies, in view of the problems of individual differences and device differences in EEG, we will continue to explore solutions with better performance and stronger generalization ability.
\end{CJK*}

\section*{Acknowledgements}
    \indent This work was supported in part by the National Natural Science Foundation of China under Grant 62176089, 62276169 and 62201356, in part by the Natural Science Foundation of Hunan Province under Grant 2023JJ20024, in part by the Key Research and Development Project of Hunan Province under Grant 2025QK3008, in part by the Key Project of Xiangjiang Laboratory under Granted 23XJ02006, in part by the STI 2030-Major Projects 2021ZD0200500, in part by the Medical-Engineering Interdisciplinary Research Foundation of Shenzhen University under Grant 2024YG008, in part by the Shenzhen University-Lingnan University Joint Research Programme, and in part by Shenzhen-Hong Kong Institute of Brain Science-Shenzhen Fundamental Research Institutions (2023SHIBS0003). 

\appendix

%\section{Appendix title 1}

%\section{Appendix title 2}


\bibliographystyle{elsarticle-num} 
\bibliography{references}



%% For example:

%% \bibitem[Aladro et al.(2015)]{Aladro15} Aladro, R., Martín, S., Riquelme, D., et al. 2015, \aas, 579, A101

%%\end{thebibliography}

\end{document}